\documentclass[dvips]{acta}
\usepackage{supertabular,lscape,epsfig}
\usepackage{amssymb}
\usepackage{amsmath}

\SetPages{431}{455}

\SetVol{59}{2009}

\begin{document}

\begin{Titlepage}
\Title{Giant Radio Sources as a probe of the cosmological evolution of the
IGM, II. The observational constraint for the model of the radio-jets propagation
through the X-ray halo--IGM interface}

\Author{E. Kuligowska, M. Jamrozy, D. Koziel-Wierzbowska, and J. Machalski}
{Astronomical Observatory, Jagiellonian University, ul. Orla 171, 30-244, Krak{\'o}w, Poland\\
e-mail:(elzbieta,jamrozy,eddie,machalsk)@oa.uj.edu.pl}

\Received{October 9, 2009}
\end{Titlepage}

\Abstract{
Three limited samples of high-redshift radio sources of FRII-type are used to constrain the
dynamical model for the jets' propagation through the two-media environment: the X-ray
emitting halo with the power-law density profile surrounding the parent galaxy and the much
hotter intergalactic medium (IGM) of a constant density. The model originally developed by
Gopal-Krishna \& Wiita (1987) is modified adopting modern values of its free parameters
taken from recent X-ray measurements with the XMM-Newton and Chandra Observatories.
We find that (i) giant-sized radio sources ($\sim$1 Mpc) exist at redshifts up to $z\sim 2$,
(ii) all newly identified the largest radio sources with $1<z<2$ appeared to be quasars,
(iii) all of them are younger and expanding faster than their counterparts at lower
redshifts, and (iv) the above properties are rather due to the powerful jets than  peculiar
environmental conditions (e.g. voids) in the IGM. The extreme powerful jets may testify
to a dominant role of the accretion processes onto black holes in earlier cosmological
epochs. 
}{Galaxies: active -- Galaxies: evolution -- intergalactic medium}

\section{Introduction}
In Paper I of this series (Machalski, Koziel-Wierzbowska \& Jamrozy, 2007) a problem
of the cosmological evolution of the intergalactic medium (IGM) was recalled and a
necessity to find distant (z$>$0.5) "giant"-sized radio sources (hereafter referred to
as GRSs) with a very low energy densities in their extended lobes, to solve this problem,
was emphasised.
Extended large-sized double radio sources are not easy to recognise because of their
relatively low radio brightness and a difficulty to detect eventual bridge connecting
brighter parts (lobes) of a common radio structure. Several observational efforts show that
most of known GRSs lie at low redshifts of z$<$0.25. For a long time this caused
a presumption that such extragalactic double radio sources, especially those of FRII-type,
did not exist at redshifts higher than about 1 because of the expected strong evolution
of a uniform IGM, $\rho_{\rm IGM}\propto (1+z)^{3}$, confining the lobes of sources (e.g.
Kapahi 1989).
The situation changed over 10 years ago when Cotter, Rawlings \& Saunders (1996) and
Cotter (1998) presented an unbiased sample of giant radio sources selected from the 7C
survey (McGilchrist et al. 1990). Their sample comprised 12 large-size sources with 
0.3$<$z$<$0.9; four of them having D$>$1 Mpc have been included in Table 1 of Paper I,
where there was shown that a list of known GRSs with z$>$0.5 and D$>$1 Mpc is very short.

The undertaken search for such GRSs on the southern sky hemisphere with the 11m SALT
telescope during the Performance Verification (P--V) phase has resulted in the detection
of 21 GRSs with the projected linear size greater than 1 Mpc.
However, we found that their redshifts do not exceed the value of 0.4 and the energy
density in only two of them is less than $10^{-14}$ J/m$^{3}$. We serendipitously
discovered that one of them (J1420$-$0545) is, in fact, the largest GRS in the Universe
(cf. Machalski et al. 2008).

Dynamical considerations relating to propagation of a bow shock at the head of powerful
supersonic jet through the galactic and/or intergalactic medium (e.g.
Arnaud et al. 1984; Begelman \& Cioffi 1989; Falle 1991) predict that the lengthening of
a channel along which the jet flows is governed by the balance between the jet's thrust
and the ram pressure of the ambient medium. This balance implies that if the channel
elongates at a rate faster than the rate at which the jet delivers energy, the motion of
the jet's head would slow down until it become subsonic. This should cause a drastic
reduction in both the speed and radiation of the head, thus effectively limiting further
growth of the radio structure both in size and luminosity.  Such a scenario is incorporated
in the analytical model for the evolution of FRII-type radio sources of Kaiser,
Dennett-Thorpe \& Alexander (1997; hereafter referred to as KDA) which combines the dynamical
model of Kaiser \& Alexander (1997) with the model for expected radio emission under the
influence of energy loss processes.

The dynamical evolution of a FRII radio source strongly depends on characteristics of the
ambient medium. Gopal-Krishna \& Wiita (1987) proposed the two-medium model consisting of
an X-ray halo around the parent galaxy with gas density decreasing with radial distance
from the galaxy and a much hotter intergalactic medium (IGM) with constant density. 
These two media were conceived to be pressure-matched at their interface. 
It is worth noting that their model (hereafter referred to as G-KW model)
does not account for the warm/hot intergalactic medium (WHIM) frequently
surrounding massive galaxies, as it was constructed before the first
assessment of the WHIM by Cen \& Ostriker (1999). The original G-KW 
model
allows a prediction of the limiting (maximum) values for the source's age and 
linear size
depending on the environment conditions, the jet power, and the cosmic epoch characterised
by the source's redshift. However, our recent detections of very large-sized radio sources
with z$>$1 and exceeding the limits predicted by their model suggests that some of its free parameters should be modified.
 
In this paper an observational constraint for the G-KW model is analysed. For this purpose, an
effort in determining the largest sizes and dynamical ages of FRII-type radio sources at 
redshifts
1$<$z$<$2 is undertaken. In Section 2, the
original G-KW model is briefly described and modified adopting modern (contemporary) values for
thermodynamic temperature and gas density of the two media. Then, the predicted relations between
the sources' linear size and the age, as well as the expansion speed of the radio lobes' 
heads
and the age are calculated. The observational data used to constrain the two-medium model are
presented in Section 3. The small sample of the most distant giant-sized radio sources (Table 1
in Paper I) is revised and supplemented with two other limited samples of FRII-type sources
comprising: (i) sources larger than 400 kpc within the redshift range 1$<$z$<$2, most of them
found in this paper, and (ii) selected 3CRR sources in majority smaller than 400 kpc at z$>$0.5
forming a comparison sample of "normal"-sized radio sources.
Physical parameters of the sample sources: the dynamical age, the jet power, the central
radio-core density and the IGM density, and others, are derived using the DYNAGE algorithm
(Machalski et al. 2007a). The application of this algorithm to the sample sources and the 
resulting values of the sources' parameters are described in Section 4. A comparison of 
the model predictions with the observational data is presented and discussed in Section 5.

For the purpose of calculating the linear size, volume and luminosity of the sample sources we
use a $\Lambda$CDM model with cosmological parameters $\Omega_{\rm m}$=0.27, $\Omega_{\rm \Lambda}$=0.73
and $H_{0}$=71 km\, s$^{-1}$Mpc$^{-1}$.

\section{The revised G-KW model}

\subsection{Base of the model}

The jet's dynamics is governed by a balance between its thrust, $\Pi_{\rm jet}\approx 
Q_{\rm jet}/v_{\rm jet}$ and the ram pressure force of the IGM, $\rho_{\rm a}v_{\rm hs}^{2}A_{\rm hs}$,
where $Q_{\rm jet}$ is the jet power, $v_{\rm jet}$
is it's velocity, $\rho_{\rm a}$ is the ambient 
medium gas density, $v_{\rm hs}$ is the speed of the jet's head
(hot spot) with which it advances into the ambient medium, and $A_{\rm hs}$ is the 
cross-sectional area of the bow shock at the end of the jet. In the G-KW model, the jet 
propagates into a two-component medium comprised of:

-- the gaseous halo with a power-law density profile $\rho_{\rm h}(d)=\rho_{0}\left[ 1+(d/a_{0})^{2}
\right]^{-\delta}$ bound to the parent optical galaxy, where $\rho_{0}$ and $a_{0}$ are the density
and the radius of the central radio core, respectively, and $\delta$=5/6. This distribution is assumed
to be invariant with redshift. It is also assumed that this halo has nearly uniform electron temperature
$(kT)_{\rm h}$[keV] (medium 1), and

-- the surrounding hotter IGM of uniform density, $\rho_{\rm IGM}$, with the temperature $(kT)_{\rm IGM}
(1+z)^{2}$[keV] (medium 2).

Similarily to Gopal-Krishna \& Wiita (1987) we have to assume characteristic values for the
density and temperature of the considered media. The values adopted hereafter for the two
components are based on the following data:

(1) The radio core radius, $a_{0}$=3 kpc is based on the fitted X-ray surface-brightness profile
of nine nearby, low-luminosity radio galaxies recently observed by Croston et al. (2008). This
value of the radius is derived from the observed angular radius of about 10 arcsec.

(2) The halos' gas temperature have been determined in a number of papers. A uniform temperature 
$(kT)_{\rm h}$=0.7 keV was measured for a few nearby, X-ray luminous elliptical galaxies with the
{\sl Chandra Observatory} by Allen et al. (2006). Using {\sl XMM-Newton} and {\sl Chandra}
observations, the values from 1 to 5 keV with a median of about 2.1 keV was found by Belsole et al.
(2007) for the X-ray clusters surrounding 20 luminous 3CRR radio sources. For the low-luminosity
radio galaxies analysed by Croston et al. (2008), a median of the fitted temperatures is about
1.4 keV. 

(3) The halos'gas (proton) density of (1 -- 2)$\times 10^{4}$ m$^{-3}$  is fitted to X-ray counts
by Belsole et al. (2007).

The interface between the X-ray halo and IGM is determined balancing the IGM pressure against the
pressure distribution in the halo. A non-relativistic gas in thermal equilibrium that has an electron
density $n_{\rm e}$[m$^{-3}$] and temperature $(kT)_{\rm e}$[keV] will have an electron pressure
$p_{\rm e}$=$n_{\rm e}\,(kT)_{\rm e}$[Pa]. Expressing electron density by the mass density,
$\rho$=$n\,\mu\,m_{\rm H}$, this balance will have place at the halo's radius $R_{\rm h}$ calculated
from

\begin{equation}
\frac{\rho_{0}}{\mu_{\rm h}m_{\rm H}}\left[1+(R_{\rm h}/a_{0})^{2}\right]^{-\delta}(kT)_{\rm h}=
\frac{\rho_{\rm IGM}}{\mu_{\rm IGM}m_{\rm H}}(kT)_{\rm IGM},
\end{equation}

\noindent
where $\mu$ and $m_{\rm H}$ are the mean molecular weight and the mass of hydrogen atom,
respectively. In this paper we assume $\mu_{\rm h}$=0.5 and $\mu_{\rm IGM}$=1.4. Besides,
for the halo (medium 1) we adopt $n_{p}$=$1.5\times 10^{4}$ m$^{-3}$, i.e. a mean proton
density of the values given by Belsole et al. (2007), which corresponds to $\rho_{0}$=$10^{-22.6}$
kg\,m$^{-3}$, and the temperature $(kT)_{\rm h}$=1.4 keV. For the IGM density we take 50\% of the
cosmic matter density, i.e. $\rho_{\rm IGM}$=$0.5\Omega_{\rm m}h^{2}\rho_{\rm clos}$=
$0.5\times 0.27\times 0.71^{2}\times (3\,H_{0}^{2})/(4\pi\,G)$, which gives  $\rho_{\rm IGM}$=
$10^{-26.9}$ kg\,m$^{-3}$. For the IGM temperature we adopt $(kT)_{\rm IGM}$=25 keV. Substituting
the above values into Eq.\,(1) we find $R_{\rm h}$=642 kpc. This radius of  X-ray halo is
compatible with the radii determined by Cassano et al. (2007) for 15 Abell cluster radio halos 
with the mean of $\sim 560\pm$170 kpc. This is worth to notice that our radius of 642 kpc is much
larger than 171 kpc used by Gopal-Krishna \& Wiita. In an expanding and uniform IGM this radius
should evolve as $R_{\rm h}(z)=642(1+z)^{-5/(2\delta)}$ kpc, i.e. $642(1+z)^{-3}$ kpc for
$\delta$=5/6.

\vspace{2mm}
Fig.1 (a), (b), (c) present the basic characteristics of the two-media model: the mass density
 $\rho(d)$, the electron temperature $kT(d)$, and the resulting electron gas pressure $p(d)$, as 
functions of the distance from the host galaxy (compact radio  core), respectively. Note that 
the balance $p_{\rm h}(R_{\rm h})=p_{\rm IGM}$ at $d=R_{\rm h}$ corresponds to a rapid 
transition between $\rho(R_{\rm h})$ and $\rho_{\rm IGM}$, as well as between $kT(R_{\rm h})$ and $kT_{\rm IGM}$, and causes an unphysical effect shown in Section 5.1. The dashed curves in Fig.1 (a) and (b) indicate desired smooth transitions between the relevant parameters which would introduce a better physical scenario into the model. 

\begin{figure}[t]
\begin{center}
\includegraphics[width=13cm]{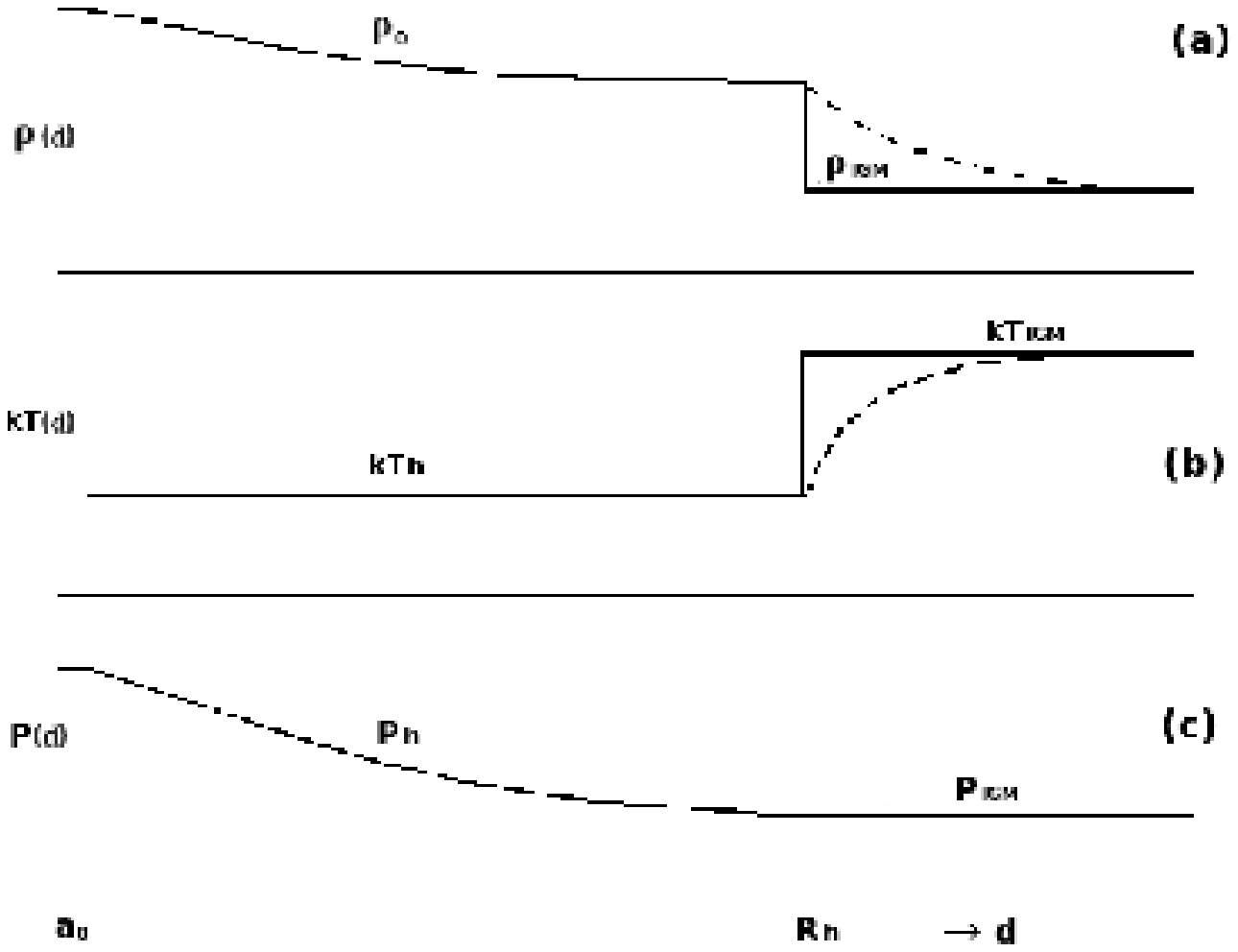}
\FigCap{Properties of the two-media environment surrounding the center of radio galaxy.}
\end{center}
\end{figure}

\vspace{2mm}

\subsection{Predictions of the model}

For $d\leq R_{\rm h}$ it is assumed that the jet propagate (through the medium 1) with a
constant opening angle, $\theta$. Under this condition, the ram pressure balance results in the
following dependence for the jet length (the radio lobe size, $D_{\ell}$) on time (the lobe's age, $t$)
and the jet's head expansion velocity, v$_{\rm h}$, on $D_{\ell}$ or $t$:

\begin{equation}
D_{\ell}(t)=\left[(2-\delta)\,A\,t\right]^{\frac{1}{2-\delta}},
\end{equation}

\begin{equation}
{\rm v}_{\rm h}(D_{\ell})=A\,D_{\ell}^{(\delta -1)},
\end{equation}

\begin{equation}
{\rm v}_{\rm h}(t)=\left[(2-\delta)\,A^{\frac{1}{\delta -1}}t\right]^{\frac{\delta -1}{2-\delta}},
\end{equation}

\noindent
where
\[A\equiv\left(\frac{4\,c_{1}Q_{\rm jet}}{\pi\theta^{2}c\,\rho_{0}a_{0}^{2\delta}}\right)^{1/2}.\]

\noindent
Here $c_{1}$ is a constant with a value between 1.5 and 3.8 depending on the source's (lobe's)
geometry described by its axial ratio $R_{\rm T}$ (cf. Kaiser \& Alexander 1997), while 
$c\approx$ v$_{\rm jet}$ is the speed of light. The jet's opening angle is also described by
$R_{\rm T}$, $\theta^{2}=c_{2}/(4\,R_{\rm T})$, where $c_{2}$ is a constant with a value between
3.6 to 4.1 depending on specific heats for the material in the jet and the lobe (cocoon),
(Eq. 17 in Kaiser \& Alexander 1997).

At $d=R_{\rm h}(z)$ the jet enters the hotter IGM (medium 2) at least an order of magnitude
less dense but pressure-matched, as shown in Fig.\,1. In order to analyse the jet's propagation
over this regime, Gopal-Krishna \& Wiita have considered two likely extreme scenarios for the
lobe's expansion: 

\vspace{2mm}
--- {\sl Scenario A} where the jet opening angle, $\theta$, is conserved. Due to a rapid
decrease of the ambient density at the interface, $\rho_{\rm IGM}\ll \rho_{\rm h}(R_{\rm h})$,
a sufficient ram-pressure will be provided only if the jet's head velocity, v$_{\rm hs}$, 
increases abruptly at $d$=$R_{\rm h}$ and then  gradually approaches the v$\propto d^{-1}$ law
expected for a constant density medium. In this scenario:

\begin{equation}
D_{\ell}(t)=\left\{2\left(K(z)+a_{0}^{\delta}\,A\left[\frac{\rho_{0}}{\rho_{\rm IGM}(1+z)^{3}}
\right]^{1/2}t\right)\right\}^{1/2}\hspace{10mm}{\rm and}
\end{equation}

\begin{equation}
{\rm v}_{\rm h}(t)=a_{0}^{\delta}\,A\left[\frac{\rho_{0}}{\rho_{\rm IGM}(1+z)^{3}}\right]^{1/2}/D_{\ell}(t),
\end{equation}

\noindent
where 
\[K(z)=\frac{1}{2}R_{\rm h}^{2}(z) - R_{\rm h}^{(2-\delta)}(z)\frac{a_{0}^{\delta}}{2-\delta}
\left[\frac{\rho_{0}}{\rho_{\rm IGM}(1+z)^{3}}\right]^{1/2}\]

\noindent
is a redshift-dependent constant providing that the time corresponding to $D_{\ell}$=$R_{\rm h}$
in Eqs.\,2 and 5 is the same.

--- {\sl Scenario B} where the jet's head velocity across the interface remains continuous
and therefore matched to the value given by Eq.\,(3) for $D_{\ell}$=$R_{\rm h}$. This can be achieved
only with an abrupt flaring of the jet's opening angle. Under this condition the model predicts:

\begin{equation}
D_{\ell}(t)=\left\{2\left(A\,R_{\rm h}^{\delta}(z)\,t-R_{\rm h}^{2}(z)\frac{\delta}{2(2-\delta)}
\right)\right\}^{1/2}\hspace{10mm}{\rm and},
\end{equation}

\begin{equation}
{\rm v}_{\rm h}(t)=A\,R_{\rm h}^{\delta}(z)/D_{\ell}(t).
\end{equation}

\noindent
The sound speed in the IGM is given by

\begin{equation}
s_{\rm IGM}(z)=\left[\frac{\Gamma(kT)_{\rm IGM}}{\mu\,m_{\rm H}}\right]^{1/2}(1+z).
\end{equation}

\noindent
Given the values $\Gamma$=5/3, $\mu$=1.4 and $(kT)_{\rm IGM}$=25 keV, the sound speed 
limiting the source (lobe) axial expansion velocity is $s_{\rm IGM}\approx 0.0056\,c$
at $z\approx 0$, and $s_{\rm IGM}\approx 0.0169\,c$ at $z\approx 2$.

The time dependence of the source size $D_{\ell}(t)$ calculated from Eqs.\,2, 5 and 7, as well
as of the axial expansion velocity v$(t)$ calculated from Eqs.\,4, 6 and 8 for the three
cases: $Q_{\rm jet}$=$10^{37.5}$ W and $z$=0.5,  $Q_{\rm jet}$=$10^{38.5}$ W and $z$=1.0,
and $Q_{\rm jet}$=$10^{39.5}$ W and $z$=2.0, are shown in Fig.~5 and Fig.~6, respectively.

To compare these predictions with the observations, we select three different samples of
FRII-type radio sources.

\section{Selection of the samples}

\subsection{Sample 1 (GRS sample)}

The revised sample of FRII-type radio sources with 0.5$<$z$<$1 and the projected linear
size larger than 1 Mpc, compiled from the literature, is presented in Table~1. Columns (1)--(5)
are self explanatory. Entries in columns (6)--(10) are the data used to determine the age
and other physical parameters of the sample sources in Section 4.
These are: $R_{\rm T}$ -- axial ratio of the source, $\phi$ -- assumed
orientation angle of the jet axis, and $P_{\nu}$ -- monochromatic luminosity
of the source at the given low, medium, and high frequency, respectively (cf.
Section 4.1). Some necessary references are given in the Notes. 

\MakeTable{lllccrccccl}{12.5cm}{The sample of FRII-type sources with 0.5$<$z$<$1 and D$>$1 Mpc. 
The radio luminosities are given in units of W/(Hz$\cdot$sr). The low-frequency luminosity, $P_{lf}$, 
is determined either at 74 MHz (marked {\sl a}) or 151 MHz ({\sl b}) or 325 MHz ({\sl c}).}
{\hline
IAU name & Survey    & z & Id.  &$D$&$R_{\rm T}$&$\phi$ & log  &log  &log  &Notes\\
         &           &   &    & [kpc] & &[$^{o}$]& $P_{lf}$ & $P_{1400}$ & $P_{4850}$ \\
 (1)     & (2)    & (3) & (4) & (5)   & (6)      & (7)  & (8) & (9) & (10) & (11)\\    
\hline
J0037$+$0027 &     & 0.5908& G & 1976 & 5.0 & 90 & 26.103$^{a}$ & 25.149 & 24.606 & (4, 5)\\
B0654$+$482  & 7C  & 0.776 & G & 1002 & 4.0 & 90 & 26.120$^{b}$ & 25.262 & 24.733 & (1)\\
J0750$+$656  &     & 0.747 & Q & 1606 & 4.0 & 90 & 26.189$^{a}$ & 25.255 & 24.715 & (3)\\
B0821$+$695  & 8C  & 0.538 & G & 2580 & 3.0 & 90 & 26.142$^{b}$ & 25.106 & 24.525 & (2)\\
B0854$+$399  & B2  & 0.528 & G & 1014 & 2.7 & 70 & 26.385$^{b}$ & 25.627 & 25.098 &\\
B1058$+$368  & 7C  & 0.750 & G & 1100 & 4.2 & 90 & 26.192$^{b}$ & 25.412 & 24.809 & (1)\\
J1130$-$1320& PKS& 0.6337& Q & 2033 & 3.8 & 90 & 27.275$^{a}$ & 26.198 & 25.613 & (4, 5)\\
B1602$+$376  & 7C  & 0.814 & G & 1376 & 2.6 & 90 & 26.292$^{b}$ & 25.486 & 24.936 & (1)\\
B1636$+$418  & 7C  & 0.867 & G & 1004 & 3.9 & 90 & 26.006$^{b}$ & 25.273 & 24.701 & (1)\\
B1834$+$620  & WNB & 0.5194& G & 1384 & 4.1 & 90 & 26.361$^{b}$ & 25.555 & 25.067 & (6)\\
J1951$+$706  &     & 0.550 & G & 1300 & 4.5 & 90 & 25.974$^{a}$ & 24.894 & 24.448 & (3)\\
J2234$-$0224&    & 0.55  & Q & 1266 & 4.5 & 90 &    ---       & 24.840 &   ---  & (4, 5)\\
\hline
\multicolumn{11}{p{12.5cm}}{
Notes: (1)--Cotter et al. (1996); (2)--Lara et al. (2000); (3)--Lara et al. (2001);
(4)--Koziel-Wierzbowska (2008); (5)--Machalski et al. (2007b); (6)-- Schoenmakers et al. (2000)}
}

\subsection{Sample 2 (Distant-source sample)}

The second sample of FRII-type sources larger than 400 kpc and having z$>$1 consists of
a few sources known from the literature and the sources found in this paper. The latter
part of this sample results from the dedicated research project attempting to determine
how large linear size FRII-type radio sources can achieve at redshifts 1$<$z$<$2.
This part was preselected using the modern Sloan Digital
Sky Survey (hereafter referred to as SDSS: Adelman-McCarthy et al. 2007) as the finding survey.
The optical objects extracted from the SDSS, fulfilling the above redshift criterion and
classified either as a galaxy (G) or a QSO (Q), were then cross-correlated with the radio
1400-MHz sky survey FIRST (Becker, White \& Helfand 1995). In the second step, all optical
objects strictly coinciding (within an angular separation less than 0.5 arcsec) with a compact
radio source, i.e. with a potential radio core, were subject to the further selection. In the
third step, we have checked whether the (core) component is surrounded by a pair of nearly
symmetric, possibly extended radio structures (lobes) with an angular separation providing
a projected linear size $D^{>}_{\sim}$400 kpc. All candidates were verified by an inspection
of their images (if exist) in other available radio surveys, namely the 1400-MHz NVSS
(Condon et al. 1998), 325-MHz WENSS (Rengelink et al. 1997), and 74-MHz VLSS (Cohen et al.
2007). The assumed cosmology (cf. Introduction) predicts that a maximum of the linear size/angular 
size quotient at z$>$1 is $\sim$8.55 kpc/arcsec which, in turn, implies that a source larger
than 400 kpc should have an angular size $LAS^{>}_{\sim}$47 arcsec. All radio 
sources selected this way from the
SDSS are classified as QSO. The final sample is presented in Table~2. All columns give the
similar data as those in Table~1. 
The radio images of exemplary sample sources, made using combined data from the NVSS and
FIRST surveys, are shown in Figs.\,2--4. The precise coordinates, angular sizes and 1400 MHz total flux
densities, taken from NVSS cataloque, are given in the Appendix (Table~8). 

However, in order to use this sample to constrain the analytical G-KW model, besides the redshift
and linear size of the member sources, we have to determine their age and other physical
parameters which is not attainable without an information about the radio spectrum within
wide-enough spectral range and providing the radio luminosity of a given source at least three
or more different observing frequencies. For the large part of the Sample~2 we were able to
use flux densities from the radio catalogues: VLSS (74 MHz), 6C and 7C (151 MHz; Hales et al.
1988 and Riley et al. 1999, respectively), WENSS (325 MHz), B3 (408 MHz; Ficarra et al. 1985,
NVSS (1400 MHz) and GB6 (4850 MHz; Gregory et al. 1996). It is worth to notice that three 
QSOs in Sample~2 have (projected) linear size larger than 1 Mpc!

\MakeTable{lllcrrccccl}{12.5cm}{The sample of FRII-type sources with 1$<$z$<$2 and D$>$400 kpc. The low-frequency luminosity marked {\sl a}, {\sl b}, {\sl c}
-- as in Table~1. The values in parenthesis are approximated because of less certain
subtraction of the core contribution}
{\hline
IAU name & Survey    & z & Id.  &$D$&$R_{\rm T}$&$\phi$ & log &  log &  log &  Notes\\
         &           &   &    & [kpc] & &[$^{o}$]&  $P_{lf}$ & $P_{1400}$ & $P_{4850}$ \\
 (1)    & (2)  & (3) & (4) & (5) & (6) & (7) & (8) & (9) & (10) & (11)\\
\hline
J0245$+$0108   &     & 1.537 & Q &  456 & 3.6 & 70 & 27.822$^{a}$ & 26.658 & 26.087 & \\
J0809$+$2015   &     & 1.129 & Q &  465 & 3.1 & 70 & ---          & (25.61)& (25.02)& 
(1)\\    
J0809$+$2912   &     & 1.481 & Q & 1120 & 3.0 & 70 & 27.234$^{a}$ & 26.476 & 26.015 & \\
J0812$+$3031   &     & 1.312 & Q & 1240 & 3.0 & 70 & 25.504$^{c}$ & 25.032 & ---    & 
(2)\\
J0819$+$0549   &     & 1.701 & Q &  985 & 3.0 & 70 & ---          & 25.578 & ---    & (1, 
2)\\
J0839$+$2928   &     & 1.136 & Q &  417 & 2.3 & 45 & 26.813$^{a}$ & 25.882 & 25.364 & \\
J0842$+$2147   &     & 1.182 & Q & 1080 & 3.0 & 70 & ---          & 25.425 & ---    & (1, 
2)\\
J0857$+$0906   &     & 1.688 & Q &  506 & 4.5 & 70 & 27.357$^{a}$ & 26.276 & 25.763 & \\
J0902$+$5707   &     & 1.595 & Q &  862 & 4.0 & 70 & 25.905$^{c}$ & 25.325 & ---    & 
(2)\\
J0906$+$0832   &     & 1.617 & Q &  682 & 3.6 & 70 & ---          & 25.750 & ---    & (1, 
2)\\
J0947$+$5154   &     & 1.063 & Q &  478 & 3.9 & 70 & 27.002$^{a}$ & 25.767 & 25.157 & \\
J0952$+$0628   &     & 1.362 & Q &  551 & 4.3 & 70 & 27.081$^{a}$ & 25.988 & 25.508 & \\
B1011$+$365    & 6C  & 1.042 & G &  416 & 3.5 & 70 & 26.902$^{b}$ & 26.204 & 25.680 & \\
J1030$+$5310   &     & 1.197 & Q &  835 & 3.0 & 90 & 26.336$^{b}$ & 25.481 & 24.945 & \\
J1039$+$0714   &     & 1.536 & Q &  501 & 3.0 & 70 & ---          & 25.450 & ---    & (1, 
2)\\
B1108$+$359    &3C252& 1.105 & G &  493 & 3.0 & 70 & 28.164$^{a}$ & 26.921 & 26.282 & \\
B1109$+$437    & B3  & 1.664 & Q &  488 & 4.5 & 70 & 28.092$^{a}$ & 27.357 & 26.812 & \\
J1130$+$3628   &     & 1.072 & Q &  422 & 3.0 & 50 & 25.992$^{b}$ & 25.277 & (24.67)& \\
J1207$-$0244 &     & 1.100 & Q &  444 & 3.0 & 50 & 26.590$^{a}$ & 25.407 & ---    & (2)\\
J1434$-$0123 &     & 1.020 & Q &  490 & 3.0 & 50 & 26.894$^{a}$ & 25.763 & ---    & (2)\\
J1550$+$3652   &     & 2.061 & Q &  675 & 3.0 & 70 & 26.771$^{b}$ & 26.041 & 25.527 & \\
J1706$+$3214   &     & 1.070 & Q &  438 & 3.0 & 50 & 26.757$^{a}$ & 25.682 & 25.150 & \\
B1723$+$510    &3C356& 1.079 & G &  614 & 4.0 & 70 & 27.875$^{b}$ & 26.916 & 26.304 & \\
J2345$-$0936 &     & 1.275 & Q &  513 & 3.2 & 90 & 27.318$^{a}$ & 26.266 & (25.62)& \\
B2352$+$796    &3C469.1&1.336& G &  626 & 4.4 & 70 & 28.269$^{a}$ & 27.167 & 26.628 & \\
\hline
\multicolumn{11}{p{12.5cm}}{
Notes: (1)--off the 7C and WENSS surveys; (2)--off the 4850-MHz GB6 or PMN6 (Griffith et 
al.
1995) surveys or below their flux density limit}
}

\begin{figure}[t]
\begin{center}
\includegraphics[width=12 cm]{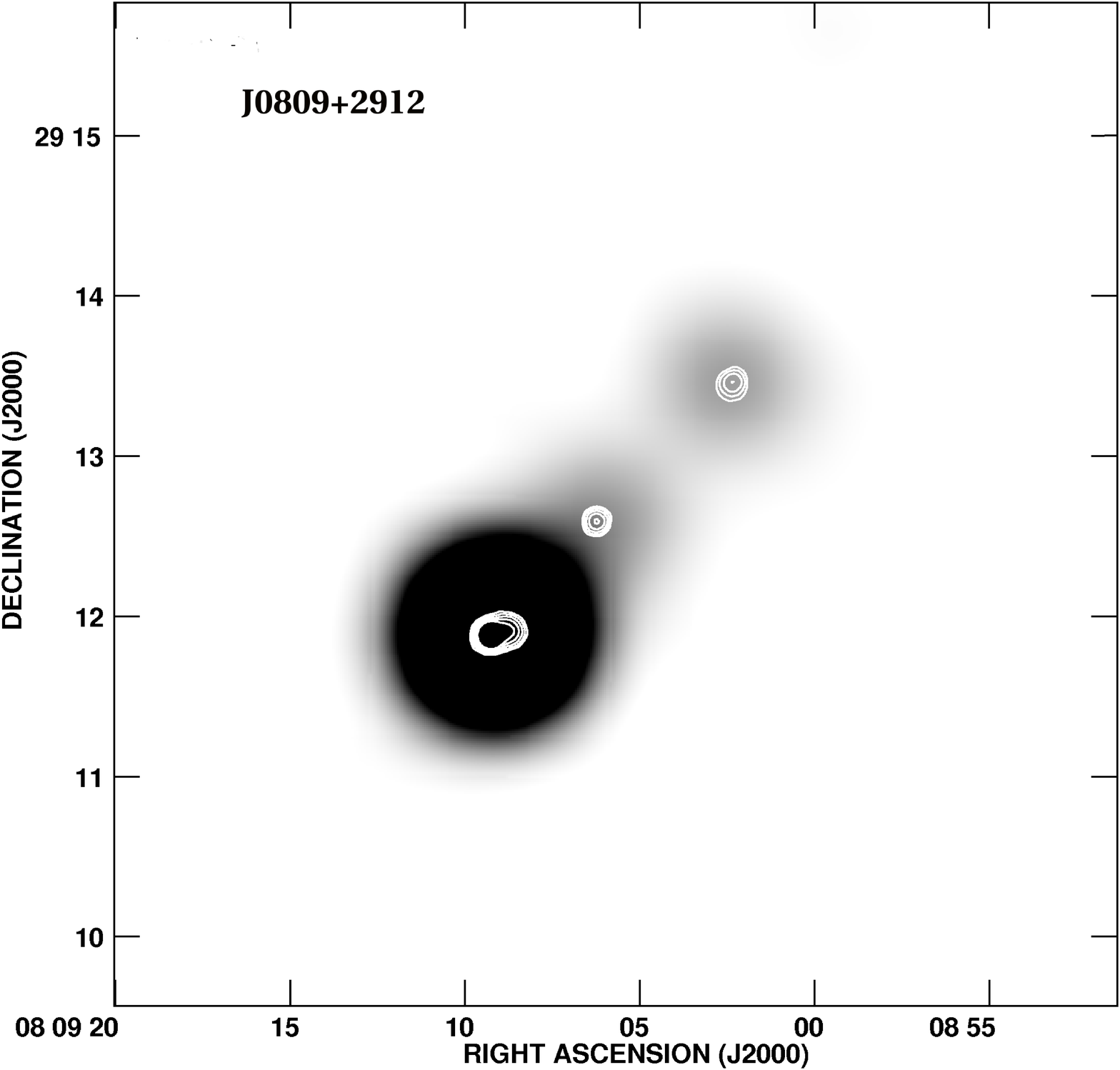}
\includegraphics[width=12 cm]{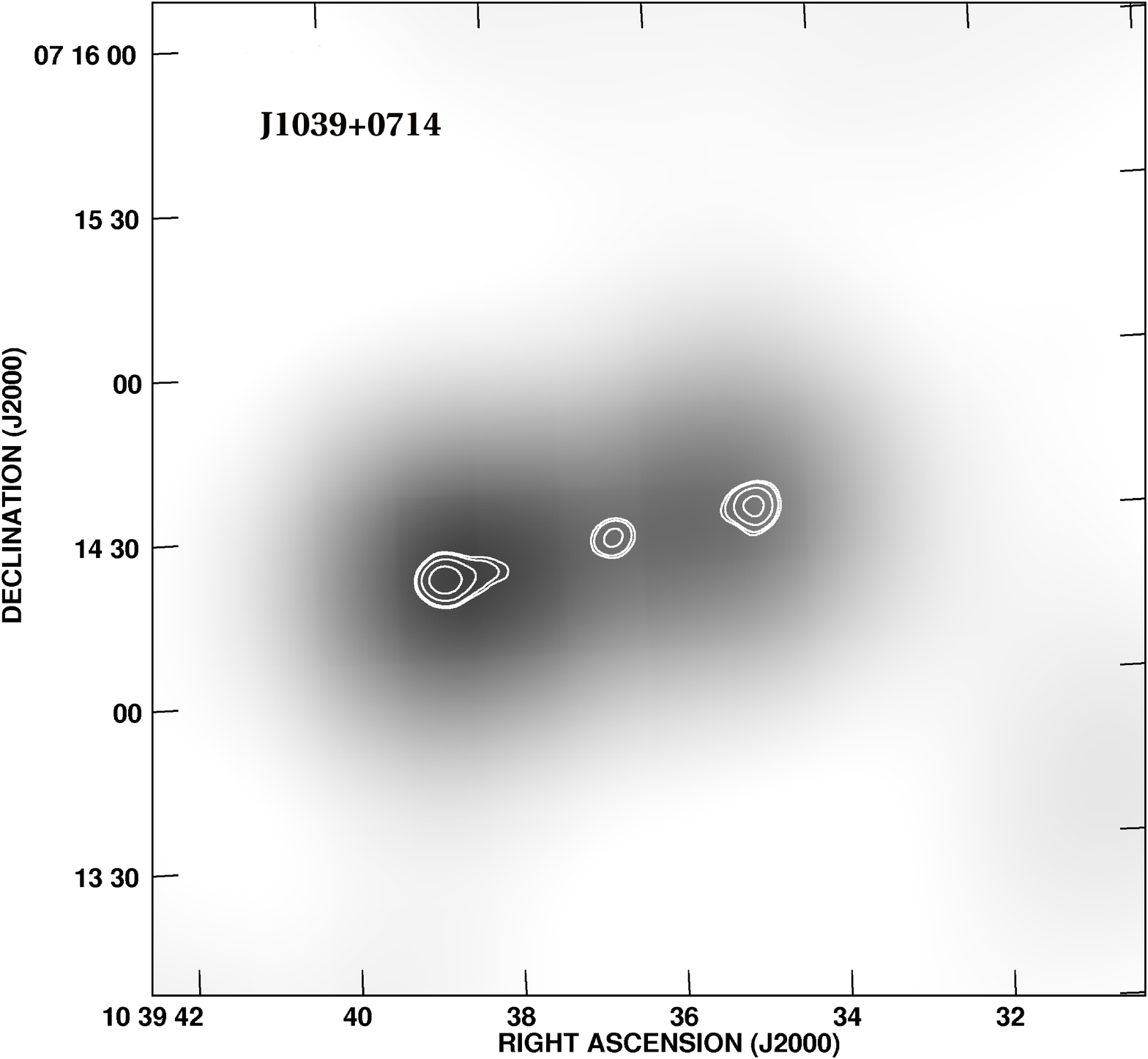}
\FigCap{1400 MHz VLA maps of the sources J0809+2912 and J1039+0714. The NVSS images (gray 
scale) are combined
with the FIRST contour maps. Total intensity, logarithmic contours are spaced by a factor of 2, starting with a value of 1.0 mJy/beam (except J1434-0123 starting with 0.2 mJy/beam).}
\end{center}
\end{figure}

\begin{figure}[t]
\begin{center}
\includegraphics[width=12 cm]{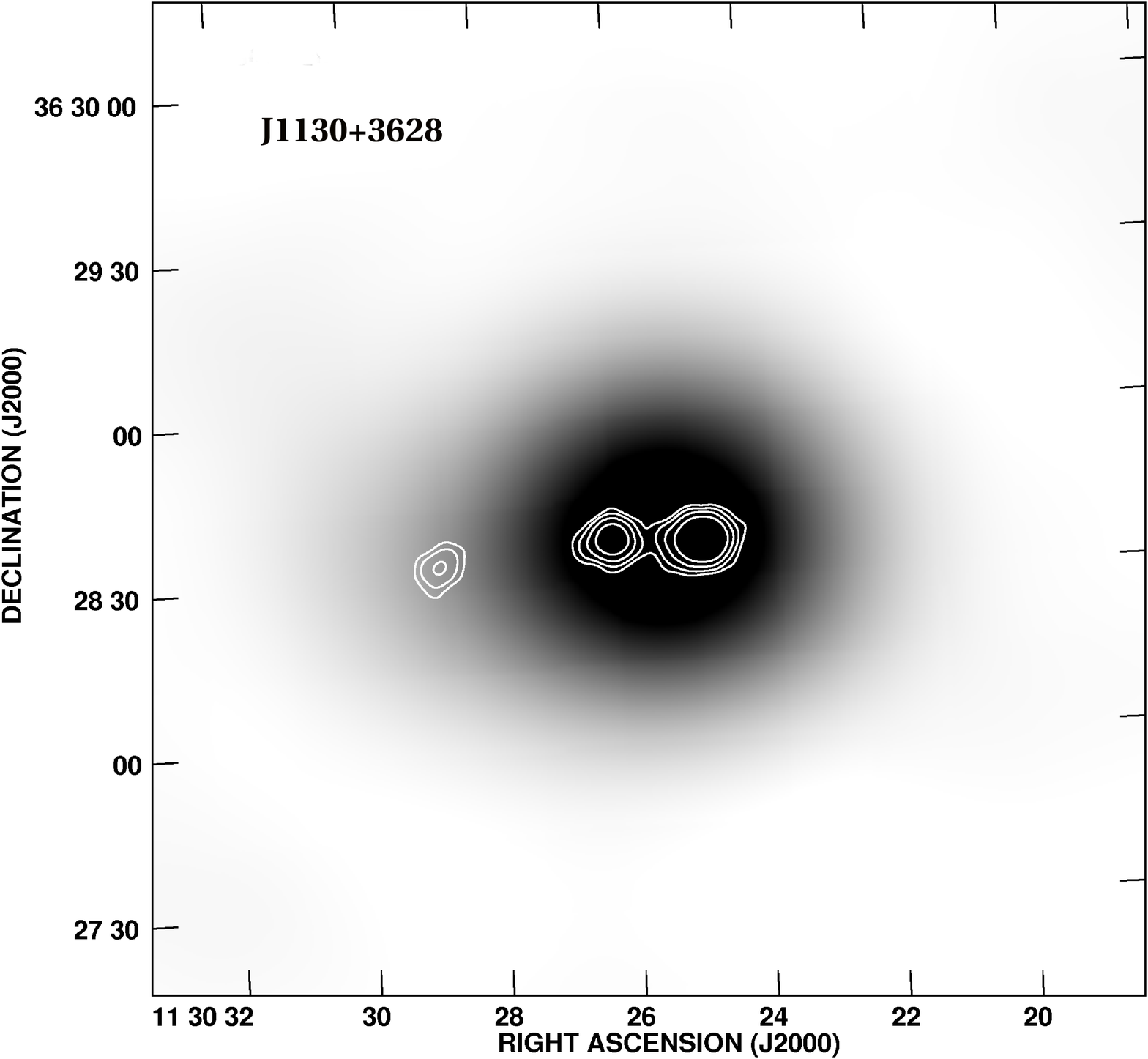}
\includegraphics[width=12 cm]{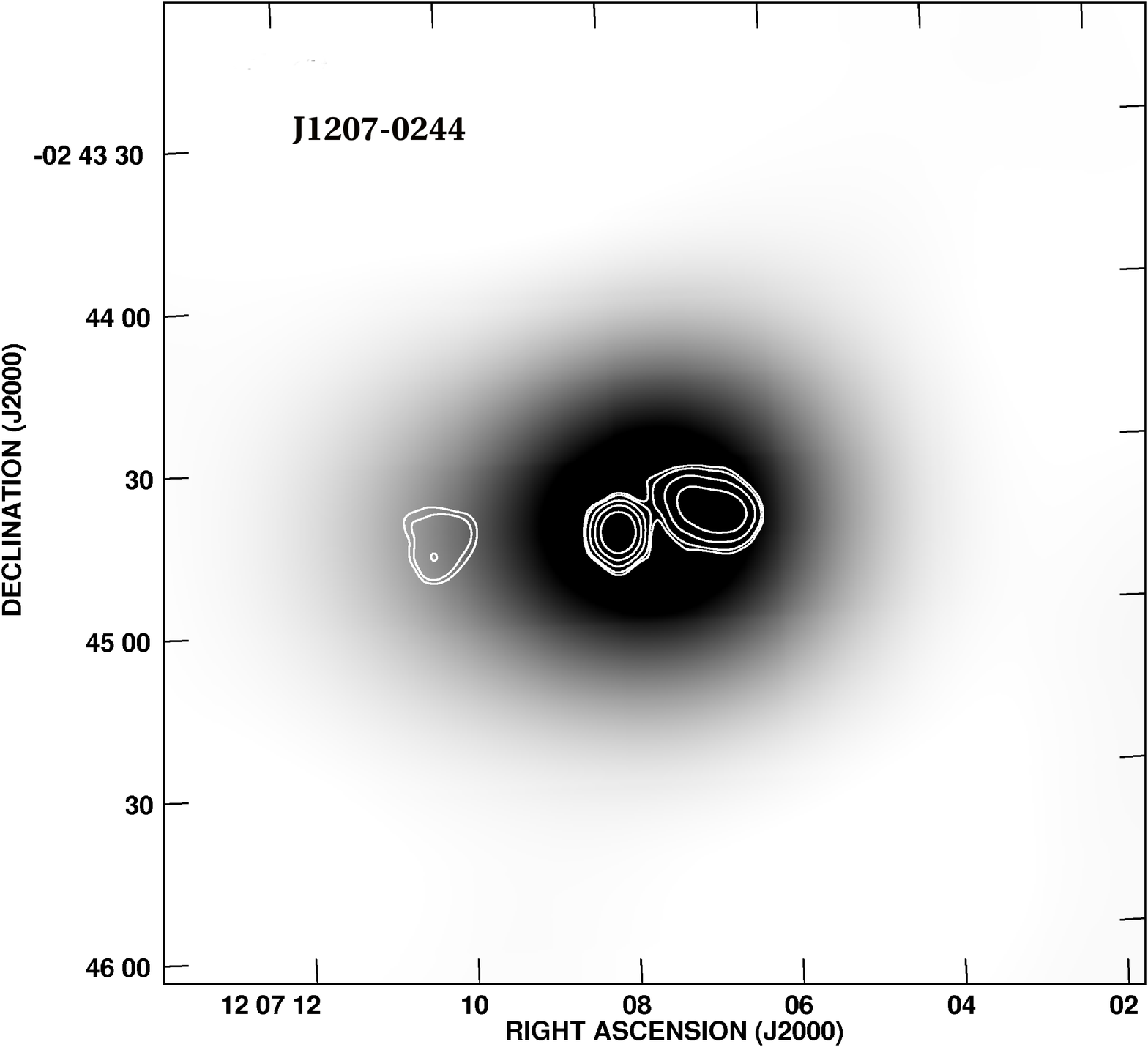}
\FigCap{The same as in Fig.2 but for J1130+3628 and J1207-0244}
\end{center}
\end{figure}

\begin{figure}[t]
\begin{center}
\includegraphics[width=12 cm]{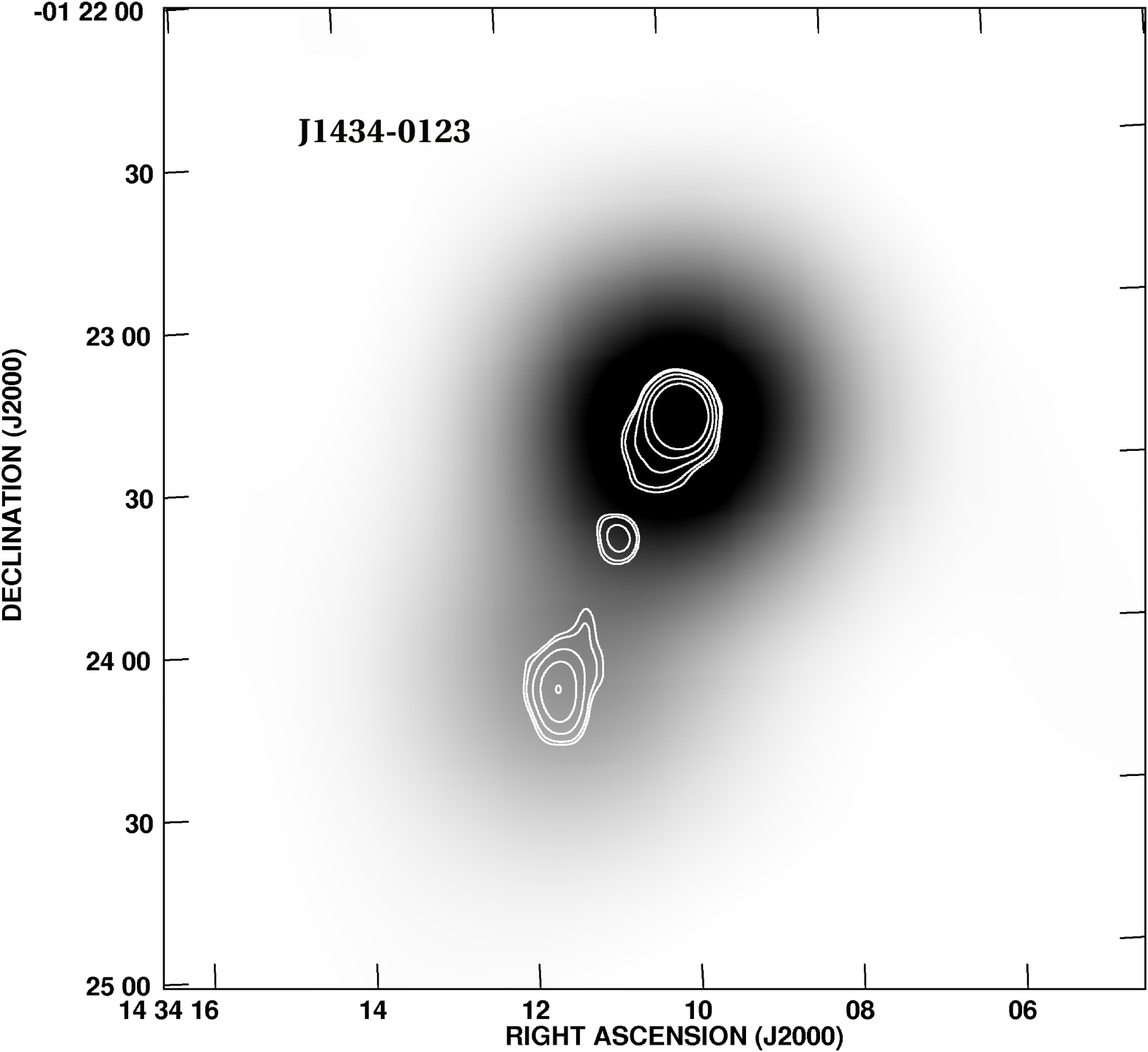}
\includegraphics[width=12 cm]{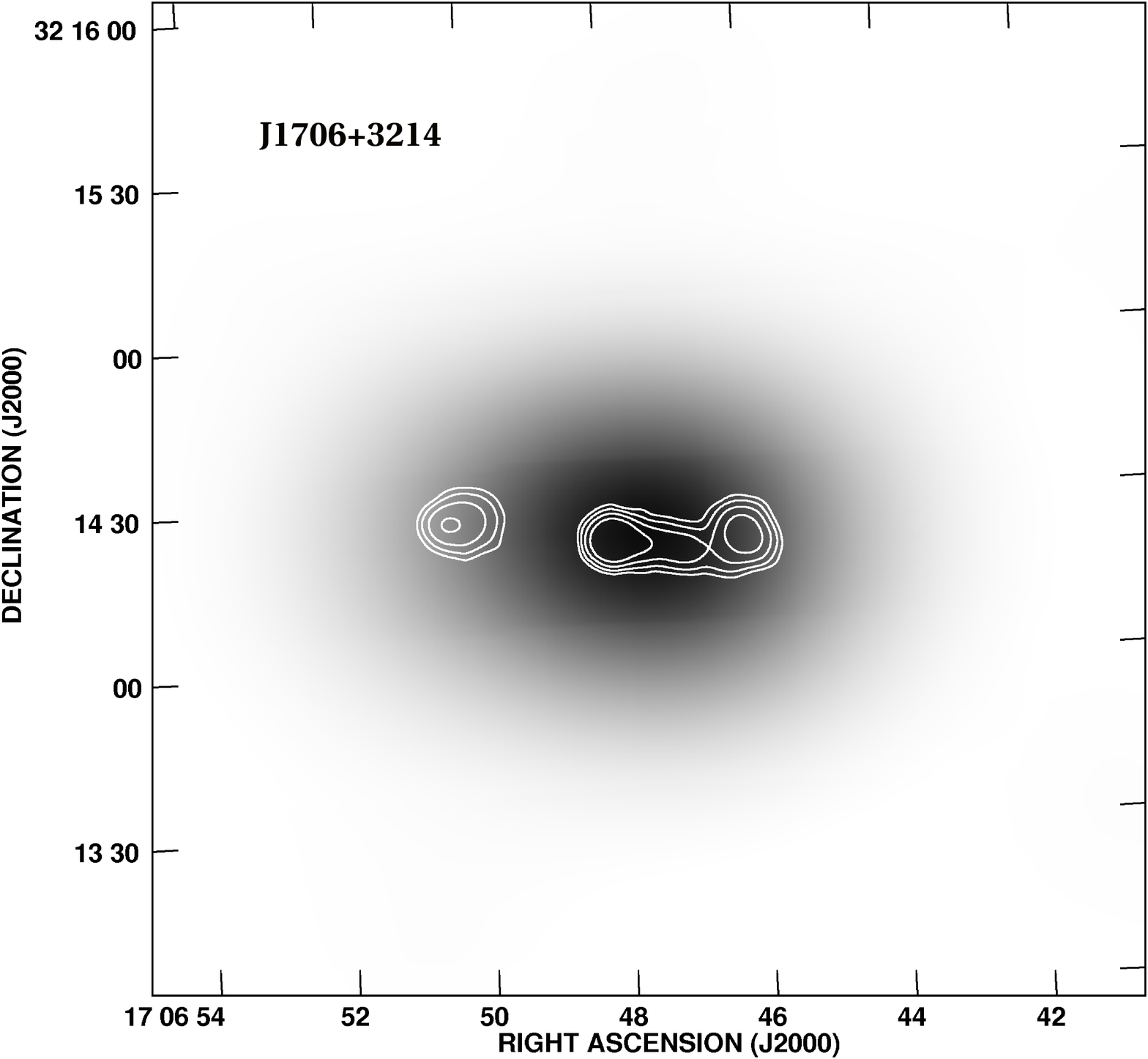}
\FigCap{The same as in Fig.2 but for J1434-0123 and J1706+3214}
\end{center}
\end{figure}

\subsection{Sample 3 (3CRR sample)}

The third observational sample comprises 3CRR FRII-type sources with z$>$0.5 but size
smaller than 400 kpc (except of 3C265 and 3C292 which are larger). For the reason to have
homogeneous and complete data for radio spectra
of the sample sources necessary for estimation of their dynamical age and corresponding
physical parameters (jet power, mean expansion velocity, etc.), this sample is limited
to sources with the Galactic latitude of $\delta_{\rm II}>30^{o}$ where good spectral data
are available from the low-frequency radio surveys: VLSS, 6C, 7C and WENSS. The sample
sources are presented in Table~3.

\MakeTable{llcrrcccc}{12.5cm}{3CRR sample of FRII-type sources with z$>$0.5 and $\delta$ > 30$^{o}$}
{\hline
Source    & z & Id.  & $D$ & $R_{\rm T}$ & $\phi$ & log & log & log \\
          &   &    & [kpc] & &[$^{o}$]&  $P_{lf}$ & $P_{1400}$ & $P_{4850}$ \\
 (1) & (2) & (3) & (4) & (5) & (6) & (7) & (8) & (9) \\
\hline
3C6.1   & 0.8404 & G & 199 & 2.7 & 70 & 27.706$^{a}$ & 26.938 & 26.494 \\
3C13    & 1.351  & G & 237 & 3.0 & 70 & 28.261$^{a}$ & 27.259 & 26.682 \\
3C22    & 0.937  & Q & 190 & 3.0 & 70 & 27.780$^{a}$ & 26.920 & 26.403 \\
3C34    & 0.689  & G & 341 & 4.3 & 90 & 27.656$^{a}$ & 26.474 & 25.815 \\
3C41    & 0.794  & G & 175 & 2.4 & 70 & 27.404$^{b}$ & 26.851 & 26.491 \\
3C54    & 0.8284 & G & 401 & 3.9 & 70 & 27.670$^{a}$ & 26.644 & 26.168 \\
3C65    & 1.174  & G & 142 & 2.3 & 70 & 28.077$^{b}$ & 27.296 & 26.761 \\
3C68.1  & 1.238  & Q & 386 & 2.4 & 50 & 27.967$^{b}$ & 27.239 & 26.767 \\
3C68.2  & 1.575  & G & 191 & 2.3 & 70 & 28.290$^{b}$ & 27.259 & 26.548 \\
3C169.1 & 0.633  & G & 315 & 2.7 & 70 & 27.287$^{a}$ & 26.214 & 25.700 \\
3C184   & 0.99   & G &  35 & 4.0 & 70 & 27.832$^{a}$ & 27.017 & 26.514 \\
3C196   & 0.871  & Q &  39 & 1.5 & 30 & 28.522$^{a}$ & 27.605 & 27.117 \\
3C204   & 1.112  & Q & 257 & 3.4 & 70 & 28.032$^{a}$ & 26.857 & 26.267 \\
3C205   & 1.534  & Q & 134 & 4.2 & 70 & 28.329$^{a}$ & 27.469 & 26.903 \\
3C217   & 0.898  & G &  86 & 4.0 & 70 & 27.796$^{a}$ & 26.853 & 26.318 \\
3C220.1 & 0.610  & G &  86 & 4.0 & 70 & 27.541$^{a}$ & 26.469 & 25.875 \\
3C220.3 & 0.685  & G &  52 & 3.0 & 70 & 27.425$^{b}$ & 26.710 & 26.045 \\
3C239   & 1.786  & G &  96 & 1.4 & 70 & 28.483$^{b}$ & 27.541 & 26.914 \\
3C247   & 0.749  & G &  95 & 2.4 & 70 & 27.350$^{b}$ & 26.746 & 26.284 \\
3C254   & 0.734  & Q &  96 & 2.0 & 50 & 27.974$^{a}$ & 26.791 & 26.198 \\
3C263   & 0.6563 & Q & 307 & 2.7 & 50 & 27.616$^{a}$ & 26.634 & 26.182 \\
3C263.1 & 0.824  & G &  41 & 1.7 & 70 & 27.696$^{b}$ & 26.911 & 26.369 \\
3C265   & 0.8108 & G & 587 & 3.9 & 70 & 27.775$^{b}$ & 26.859 & 26.310 \\
3C266   & 1.275  & G &  36 & 2.4 & 70 & 27.981$^{b}$ & 27.137 & 26.509 \\
3C268.4 & 1.400  & Q &  83 & 2.8 & 50 & 28.123$^{a}$ & 27.273 & 26.774 \\
3C270.1 & 1.519  & Q & 102 & 1.3 & 50 & 28.314$^{a}$ & 27.478 & 26.917 \\
3C272   & 0.944  & G & 461 & 3.3 & 70 & 27.417$^{b}$ & 26.708 & 26.135 \\
3C280   & 0.996  & G & 113 & 1.5 & 70 & 27.990$^{b}$ & 27.288 & 26.829 \\
3C280.1 & 1.659  & Q & 169 & 4.0 & 70 & 28.267$^{a}$ & 27.343 & 26.771 \\
3C289   & 0.967  & G &  81 & 1.3 & 70 & 27.686$^{b}$ & 26.966 & 26.430 \\
3C292   & 0.71   & G & 960 & 4.2 & 90 & 27.406$^{b}$ & 26.547 & 26.075 \\
3C294   & 1.779  & G & 135 & 1.9 & 70 & 28.413$^{b}$ & 27.431 & 26.840 \\
3C322   & 1.681  & G & 283 & 2.5 & 70 & 28.171$^{b}$ & 27.485 & 26.954 \\
3C324   & 1.207  & G &  85 & 1.8 & 70 & 27.999$^{b}$ & 27.257 & 26.712 \\
3C325   & 0.860  & G & 122 & 2.2 & 70 & 27.657$^{b}$ & 27.011 & 26.481 \\
3C330   & 0.549  & G & 395 & 5.4 & 90 & 27.432$^{b}$ & 26.771 & 26.334 \\
3C337   & 0.635  & G & 295 & 2.5 & 70 & 27.522$^{a}$ & 26.599 & 26.125 \\
3C352   & 0.8057 & G &  75 & 2.3 & 70 & 27.695$^{a}$ & 26.699 & 26.108 \\
3C427.1 & 0.572  & G & 153 & 2.2 & 70 & 27.437$^{b}$ & 26.634 & 26.071 \\
3C437   & 1.48   & G & 316 & 3.9 & 70 & 28.086$^{b}$ & 27.481 & 27.030 \\
3C441   & 0.707  & G & 236 & 2.5 & 70 & 27.395$^{b}$ & 26.636 & 26.166 \\
3C470   & 1.653  & G & 205 & 4.0 & 70 & 28.269$^{a}$ & 27.444 & 26.952 \\
\hline
}

\section{Ageing analysis of the samples' sources}

\subsection{Application of the DYNAGE algorithm}

The dynamical age analysis is performed using the DYNAGE algorithm of Machalski
et al. (2007a). It is based on the analytical KDA model (cf. Introduction).
The original KDA model, assuming values for a number of free parameters of the
model (e.g. the jet power $Q_{\rm jet}$, radius $a_{0}$ and density $\rho_{0}$
of the radio core, exponent $\beta$ in the power-law density distribution in
the external medium $\rho(d/a_{0})^{-\beta}$, and the effective injection
spectral index $\alpha_{\rm inj}$, which approximates the initial relativistic
particle continuum at a head of the jet $n(\gamma_{i})=n_{0}(\gamma_{i}^{-p})$,
where $\gamma_{i}$ is the Lorentz factor of these particles and
$p=(\alpha_{\rm inj}+1)/2)$, allows prediction of a time evolution of the
source's parameters, e.g. its length, $D(t)$, and radio luminosity,
$P_{\nu}(t)$, at a given observing frequency -- as a function of its assumed
age $t$.

Oppositely, the DYNAGE algorithm enables  solving a reverse problem, i.e.
finding the values of $t$, $\alpha_{\rm inj}$, $Q_{\rm jet}$, and $\rho_{0}$
for a real source. Determining the values of these four free parameters of the
model is possible by a fit to the observational parameters of a source: its
projected linear size $D$, the volume $V$, the radio monochromatic luminosity
$P_{\nu}$, and the radio spectrum $\alpha_{\nu}$ that provides $P_{\nu,i}$ at
a number of observing frequencies $i=1, 2, 3,...$  To determine these
luminosities for the sample sources, we have had to fix the flux density in
their lobes at a number of frequencies, i.e. to subtract a flux contribution
from the compact components such as the radio core and hot spots in these
lobes. This was especially important for the sample sources identified with
quasars.  The resulting luminosities at three observing frequencies: the lowest
frequency used (74, 151, or 325 MHz), the medium one of 1400 MHz, and the
highest one of 4850 MHz, are given in columns (8)--(10) in Tables 1 and 2, as
well as in columns (7)--(9) in Table~3.

As in KDA, we adopted a cylindrical geometry of the source's cocoon (radio
lobes) where its volume is determined by the deprojected length $D/\sin\phi$
and the axial ratio $R_{\rm T}$. The values of $R_{\rm T}$ are estimated from
the low-frequency radio images, and an angle of orientation of the jet axis to
the observer's line of sight, $\phi$, is subjectively estimated from the
observed asymmetry in the lobes' arms and brightness. The valus of these two
parameters are given in columns (6) and (7) of Tables 1 and 2, and in columns
(5) and (6) in Table~3.

Unfortunately, the values of other free parameters of the model have to be assumed. These  
are: $\gamma_{i,{\rm min}}=1$ and $\gamma_{i,{\rm max}}=10^{7}$
that are the Lorentz factors determining the energy range of the relativistic particles 
used in integration of their initial power-law distribution; $\Gamma_{\rm j}$, 
$\Gamma_{\rm x}$,
$\Gamma_{\rm B}$, and $\Gamma_{\rm c}$, are all equal to 5/3, are the adiabatic indices in the equation of state
for the jet material, the unshocked medium surrounding the lobes, a "magnetic" fluid, and
the source (cocoon) as a whole, respectively; and $k^{\prime}=0$ is the ratio of the 
energy density of thermal particles to that of the relativistic particles. The more 
detailed description of application of the above algorithm to observed radio structures is 
published in Machalski et al. (2009).

\subsection{Results of the modelling}

The resulting values of the age and other physical parameters of the sample sources are
given in Tables 4, 5 and 6 (for the Sample 1, 2 and 3, respectively). Columns (2) and
(3) give the dynamical age, $t$, and the effective, initial slope of the radio spectrum
$\alpha_{\rm inj}$. Columns (4) -- (7): (in the logarithmic scale) the jet power, $Q_{\rm jet}$,
the central core density, $\rho_{0}$, the pressure along the jet axis, $p_{\rm h}$, and the
total energy radiated out during the age of source, $U_{\rm c}$. The age $t$ and the fitted
values of $Q_{\rm jet}$ and  $\rho_{0}$ fulfil the dynamical equation

\begin{equation}
t=\left(\frac{D}{2\,c_{1}\sin\phi}\right)^{\frac{5-\beta}{3}}\left(\frac{\rho_{0}a_{0}^{\beta}}
{Q_{\rm jet}}\right)^{1/3},
\end{equation}

\noindent
Where $D/(2\,\sin \phi)\equiv D_{\ell}$ is the deprojected lenght of a given lobe treated 
here as one 
half of the total, observed size of the source, $D$. The energy density fulfilling the energy equipartition condition is calculated from

\begin{equation}
u_{\rm eq}(t)=\frac{18\,c_{1}^{(2-\beta)}}{(\Gamma_{\rm x}+1)(\Gamma_{\rm c}-1)(5-\beta)^{2}
{\cal P}_{\rm hc}}\left(\rho_{0}a_{0}^{\beta}\right)^{\frac{3}{5-\beta}}
Q_{\rm jet}^{\frac{2-\beta}{5-\beta}}t^{-\frac{4+\beta}{5-\beta}},
\end{equation}

\noindent
where ${\cal P}_{\rm hc}$=$(2.14-0.52\beta)R_{\rm T}^{2.04-0.25\beta}$ is the empirical formula
for the pressure ratio along the jet axis and the transverse direction taken from Kaiser (2000).
The total radiated energy is simply $U_{\rm c}$=$u_{\rm eq}\times V_{\rm c}$, where the source
(cocoon) volume attained at the age $t$ is $V_{\rm c}$=$(\pi/4)\,D^{3}/(2R_{\rm T})^{2}$.
Column (8) gives the magnetic field strength estimate derived from

\begin{equation}
B=\left(\frac{24}{7}10^{11}\pi\,u_{\rm B}\right)^{1/2}\hspace{2mm}[{\rm nT}],
\end{equation}

\noindent
where the magnetic field energy density is $u_{\rm B}$=$u_{\rm eq}(1+p)/(5+p)$. The last
column (9) gives the actual expansion velocity of the lobe's head along the jet axis which is the derivative
of the $D_{\ell}$ function

\begin{equation}
{\rm v}_{\rm h}(t)=\frac{dD_{\ell}}{dt}=\frac{3\,c_{1}}{5-\beta}\left(\frac{Q_{\rm 
jet}}{\rho_{0}a_{0}^{\beta}}
\right)^{\frac{1}{5-\beta}}t^{\frac{\beta -2}{5-\beta}}.
\end{equation}

\section{Observational constrain for the model}
\subsection{Comparison of the model's prediction with the observational data}

The model's predictions, i.e. the dependence of the deprojected linear size and the jet's head 
velocity on the source's age (${D}/{sin\phi}$  vs $t$ and $v_{\rm h}/c$ vs $t$, 
respectively), are compared with the data determined for the sampled sources and given in 
Tables 4, 5 and 6. Fig. 5 shows the dependence of D/sin$\phi$ on $t$. The model's 
predictions are presented for three different sets of values of the jet power and redshift: 
$z=0.5$ and $Q_{\rm jet}=10^{37.5}$ W, $z=1$ and $Q_{\rm jet}=10^{38.5}$ W, $z = 2$ and 
$Q_{\rm jet}=10^{39.5}$ W, marked "1", "2", and "3", respectively. The above dependence 
predicted in the frame of scenario A are drawn with the solid lines, while these for scenario B 
-- with the dashed lines. The points of bifurcation of the model's predictions into scenario A
 and scenario B correspond to the halo diameter dependent on redshift, $2R_{\rm h}(z)$ (on the ordinate axis) 
and to the related age (on the abscissa axis). The ends of both solid and dashed lines 
indicate a maximum size and age at which the predicted expansion velocity reaches the speed of sound. The crosses show the distribution of the GRSs from Sample 1, 
the full circles -- the distant sources (mostly QSOs) from Sample 2, and the open circles -- the 3CRR sources from Sample 3.

The dependence of v$_{\rm h}/c$  on $t$ is shown in Fig. 6. The model's predictions and the sample sources are indicated with the lines and symbols as in Fig. 5. The rapid increase of v$_{\rm h}/c$ in the frame of scenario A results from the discontinuities of ${\rho(d)}$ and $kT(d)$ at the halo-IGM interface (cf. Fig. 1). The three horizontal dotted lines indicate the predicted lower values of the axial expansion velocity,
v$_{\rm h,min}/c$, limited by the sound speed at the given values of redshift. 

The maximum size which any FRII-type radio source can reach with a given value of v$_{\rm h,min}/c$ , (i.e. at the corresponding redshift) depends on the jet power. The size $D_{\rm max}$ in the frame of scenario A, as a function of v$_{\rm h,min}/c$ (or $z$), is plotted on Fig. 7 for three values of $Q_{\rm jet}$ considered above.

\MakeTable{lclcccrccccll}{12.5cm}{Age and physical parameters of the sources in Sample~1}
{\hline
IAU name &&t&$\alpha_{\rm inj}$&log\,$Q_{\rm jet}$&log\,$\rho_{0}$&log\,$p_{\rm h}$ &log\,U$_{\rm c}$&
B$_{\rm eq}$ & v$_{\rm h}$/c \\
         &&[Myr]&    &  [W] & [kg/m$^{3}$] & [N/m$^{2}$] & [J] & [nT] \\
 (1) & & (2) & (3) & (4) & (5) & (6) & (7) & (8) & (9)\\
\hline
J0037+0027 &&$\;\:$59.0 & 0.506 & 38.474 & $-$23.311 & $-$12.13 & 53.00 & 0.16 & 0.047 \\
B0654+482  &&$\;\:$57.0 & 0.555 & 38.296 & $-$22.741 & $-$11.68 & 52.92 & 0.33 & 0.024 \\
J0750+656  &&$\;\:$39.0 & 0.507 & 38.563 & $-$23.670 & $-$12.20 & 53.02 & 0.18 & 0.057 \\
B0821+695  && 135.0 & 0.645 & 38.710 & $-$22.968 & $-$12.45 & 53.84 & 0.17 & 0.027 \\
B0854+399  &&$\;\:$69.0 & 0.522 & 38.362 & $-$22.992 & $-$12.04 & 53.24 & 0.29 & 0.023 \\
B1058+368  &&$\;\:$57.0 & 0.545 & 38.384 & $-$22.741 & $-$11.66 & 52.98 & 0.32 & 0.027 \\
J1130$-$1320&& 100.0& 0.537 & 39.005 & $-$22.431 & $-$11.71 & 53.90 & 0.33 & 0.028 \\
B1602+376  &&$\;\:$47.0 & 0.524 & 38.621 & $-$23.648 & $-$12.35 & 53.35 & 0.22 & 0.041 \\
B1636+418  &&$\;\:$38.0 & 0.526 & 38.367 & $-$23.229 & $-$11.82 & 52.83 & 0.29 & 0.037 \\
B1834+620  &&$\;\:$75.0 & 0.533 & 38.455 & $-$22.688 & $-$11.80 & 53.18 & 0.28 & 0.026 \\
J1951+706  &&$\;\:$73.0 & 0.525 & 38.061 & $-$22.921 & $-$12.02 & 52.73 & 0.25 & 0.025 \\
J2234$-$0224&&$\;$(91.0)&      & (38.1) &($-$22.55) & & & & 0.019 \\
\hline
}

\MakeTable{lllccrccccl}{12.5cm}{Age and physical parameters of the sources in Sample~2}
{\hline
IAU name &&t&$\alpha_{\rm inj}$&log\,$Q_{\rm jet}$&log\,$\rho_{0}$&log\,$p_{\rm h}$ &log\,U$_{\rm c}$&
B$_{\rm eq}$ & v$_{\rm h}$/c \\
         &&[Myr]&    &  [W] & [kg/m$^{3}$] & [N/m$^{2}$] & [J] & [nT] \\
(1) & & (2) & (3) & (4) & (5) & (6) & (7) & (8) & (9)\\
\hline  
J0245+0108   && 18.5 & 0.603 & 39.295 & $-$22.153 & $-$10.30 & 53.45 & 1.74 & 0.038 \\    
J0809+2912   &&$\;\:$7.5 & 0.496 & 39.715 & $-$24.555 & $-$11.65 & 53.59 & 0.41 & 0.233 \\
J0839+2928   && 21.5 & 0.535 & 38.655 & $-$23.031 & $-$11.44 & 52.93 & 0.62 & 0.031 \\
J0857+0906   && 14.6 & 0.564 & 39.111 & $-$22.629 & $-$10.48 & 53.08 & 1.13 & 0.054 \\
J0947+5154   && 50.5 & 0.624 & 38.508 & $-$21.616 & $-$10.62 & 53.06 & 1.12 & 0.015 \\
J0952+0628   && 23.0 & 0.554 & 38.806 & $-$22.453 & $-$10.75 & 52.98 & 0.89 & 0.037 \\
B1011+365    && 18.1 & 0.542 & 38.752 & $-$22.686 & $-$10.76 & 52.94 & 1.00 & 0.036 \\
J1030+5310   && 28.0 & 0.548 & 38.605 & $-$23.408 & $-$11.77 & 53.05 & 0.38 & 0.042 \\
B1108+359    && 23.3 & 0.655 & 39.560 & $-$21.913 & $-$10.22 & 53.90 & 2.18 & 0.031 \\
B1109+437    &&$\;\:$4.3 & 0.544 & 40.028 & $-$23.238 & $-$10.04 & 53.47 & 1.87 & 0.175 \\
J1130+3628   && 29.5 & 0.536 & 38.180 & $-$22.898 & $-$11.39 & 52.53 & 0.47 & 0.027 \\
J1550+3652   &&$\;\:$6.6 & 0.511 & 39.340 & $-$24.327 & $-$11.42 & 53.16 & 0.53 & 0.160 \\
J1706+3214   && 36.6 & 0.572 & 38.460 & $-$22.340 & $-$11.00 & 52.90 & 0.72 & 0.022 \\
B1723+510    && 23.5 & 0.661 & 39.676 & $-$21.881 & $-$10.09 & 53.92 & 1.98 & 0.041 \\
J2345$-$0936 && 29.6 & 0.564 & 38.797 & $-$22.308 & $-$10.82 & 53.23 & 1.05 & 0.024 \\
B2352+796    && 16.0 & 0.578 & 39.747 & $-$22.155 & $-$10.08 & 53.74 & 1.83 & 0.058 \\
\hline
}
   
\MakeTable{lccccrcrc}{12.5cm}{Age and physical parameters of the sources in Sample~3.The age solution for the sources which name is followed by the asterisk is provided by J. Machalski (unpublished).} 
{\hline
IAU name &t&$\alpha_{\rm inj}$&log\,$Q_{\rm jet}$&log\,$\rho_{0}$&log\,p$_{\rm h}$&log\,U$_{\rm c}$&
B$_{\rm eq}$ & v$_{\rm h}$/c \\
         &[Myr]&    &  [W] & [kg/m$^{3}$] & [N/m$^{2}$] & [J] & [nT] \\
(1) &  (2) & (3) & (4) & (5) & (6) & (7) & (8) & (9)\\
\hline 
3C6.1   &$\;\;$ 5.12 & 0.564 & 39.298 & $-$22.891 & $-$10.08 & 53.03 & 2.75 & 0.057 \\
3C13    &$\;\;$ 5.87 & 0.583 & 39.606 & $-$22.549 & $-$9.81 & 53.35 & 3.41 & 0.060 \\
3C22    &$\;\;$ 4.68 & 0.559 & 39.316 & $-$22.799 & $-$9.92 & 52.96 & 3.03 & 0.060 \\
3C34    & 24.0 & 0.630 & 39.012 & $-$21.434 & $-$9.86 & 53.22 & 2.56 & 0.018 \\
3C41    &$\;\;$ 1.52 & 0.534 & 39.580 & $-$24.250 & $-$10.34 & 52.85 & 2.21 & 0.182 \\
3C54*    & 15.2 & 0.562 & 39.141 & $-$22.280 & $-$10.26 & 53.17 & 1.64 & 0.039 \\
3C65    &$\;\;$ 3.40 & 0.588 & 39.604 & $-$22.881 & $-$9.74 & 53.23 & 4.65 & 0.065 \\
3C68.1*  &$\;\;$ 7.94 & 0.564 & 39.784 & $-$23.053 & $-$10.43 & 53.66 & 1.70 & 0.089 \\
3C68.2  &$\;\;$ 4.35 & 0.771 & 40.209 & $-$22.406 & $-$9.41 & 53.94 & 6.97 & 0.076 \\
3C169.1 & 24.7 & 0.575 & 38.611 & $-$22.230 & $-$10.68 & 53.02 & 1.37 & 0.019 \\
3C184   &$\;\;$ 0.37 & 0.577 & 39.471 & $-$23.395 &  $-$8.46 & 52.01 & 12.83& 0.142 \\
3C196   &$\;\;$ 0.60 & 0.579 & 40.042 & $-$23.374 &  $-$8.91 & 52.82 & 10.14& 0.188 \\
3C204   & 11.6 & 0.625 & 39.350 & $-$21.969 & $-$9.76 & 53.36 & 3.28 & 0.035 \\
3C205   &$\;\;$ 1.45 & 0.584 & 39.949 & $-$22.840 & $-$8.98 & 52.95 & 6.76 & 0.143 \\
3C217   &$\;\;$ 1.87 & 0.588 & 39.301 & $-$22.481 &  $-$8.97 & 52.41 & 7.08 & 0.068 \\
3C220.1 & 11.0 & 0.610 & 38.900 & $-$21.653 & $-$9.53 & 52.78 & 3.74 & 0.024 \\
3C220.3 &$\;\;$ 1.92 & 0.580 & 38.955 & $-$22.345 &  $-$8.97 & 52.21 & 9.01 & 0.040 \\
3C239*   &$\;\;$ 2.90 & 0.673 & 39.903 & $-$22.904 &  $-$9.68 & 53.60 & 7.61 & 0.052 \\
3C247*   &$\;\;$ 1.70 & 0.547 & 39.182 & $-$23.182 & $-$9.88 & 52.50 & 3.77 & 0.087 \\
3C254   &$\;\;$ 5.60 & 0.627 & 39.172 & $-$22.240 & $-$9.60 & 52.98 & 5.18 & 0.031 \\
3C263   & 12.8 & 0.557 & 39.126 & $-$22.584 & $-$10.44 & 53.15 & 1.53 & 0.044 \\
3C263.1* &$\;\;$ 1.30 & 0.600 & 39.133 & $-$23.110 &  $-$9.40 & 52.43 & 8.83 & 0.050 \\
3C265*   & 27.0 & 0.609 & 39.415 & $-$21.905 & $-$10.25 & 53.72 & 1.66 & 0.034 \\
3C266*   &$\;\;$ 0.86 & 0.639 & 39.462 & $-$22.674 &  $-$8.64 & 52.48 & 16.12& 0.065 \\
3C268.4 &$\;\;$ 0.95 & 0.574 & 39.809 & $-$23.280 & $-$9.15 & 52.70 & 6.56 & 0.159 \\
3C270.1* &$\;\;$ 2.35 & 0.568 & 39.708 & $-$23.458 & $-$10.07 & 53.26 & 4.30 & 0.079 \\
3C272*   & 19.0 & 0.557 & 39.132 & $-$22.484 & $-$10.57 & 53.37 & 1.32 & 0.038 \\
3C280*   &$\;\;$ 2.60 & 0.560 & 39.520 & $-$23.578 & $-$10.23 & 53.16 & 3.76 & 0.068 \\
3C280.1 &$\;\;$ 2.50 & 0.574 & 39.795 & $-$22.635 & $-$9.23 & 53.03 & 5.24 & 0.100 \\
3C289*   &$\;\;$ 3.60 & 0.571 & 39.080 & $-$23.296 & $-$10.30 & 52.88 & 3.92 & 0.035 \\
3C292*   & 38.0 & 0.566 & 39.180 & $-$22.266 & $-$10.86 & 53.60 & 0.81 & 0.035 \\
3C294*   &$\;\;$ 3.65 & 0.683 & 39.950 & $-$22.611 & $-$9.54 & 53.67 & 6.95 & 0.058 \\
3C322*   &$\;\;$ 4.00 & 0.557 & 39.908 & $-$23.310 & $-$10.16 & 53.58 & 2.66 & 0.111 \\
3C324*   &$\;\;$ 2.20 & 0.584 & 39.471 & $-$23.117 & $-$9.71 & 52.98 & 5.91 & 0.061 \\
3C325*   &$\;\;$ 2.90 & 0.566 & 39.320 & $-$23.197 & $-$9.95 & 52.89 & 3.77 & 0.066 \\
3C330*   &$\;\;$ 6.60 & 0.543 & 39.404 & $-$22.699 & $-$9.98 & 52.93 & 1.82 & 0.084 \\
3C337   & 11.3 & 0.553 & 38.999 & $-$22.853 & $-$10.64 & 53.10 & 1.53 & 0.039 \\
3C352   &$\;\;$ 2.96 & 0.595 & 38.986 & $-$22.631 & $-$9.55 & 52.53 & 5.78 & 0.038 \\
3C427.1* &$\;\;$ 9.10 & 0.592 & 38.903 & $-$22.498 & $-$10.18 & 52.98 & 2.91 & 0.026 \\
3C437*   &$\;\;$ 2.42 & 0.546 & 40.113 & $-$23.405 & $-$9.80 & 53.37 & 2.78 & 0.205 \\
3C441*   & 10.0 & 0.556 & 38.984 & $-$22.764 & $-$10.45 & 53.05 & 1.91 & 0.037 \\
3C470   &$\;\;$ 1.97 & 0.550 & 39.973 & $-$23.063 & $-$9.41 & 53.10 & 4.25 & 0.155 \\
\hline 
}

\subsection{Age and physical parameters of the sample sources}

The three samples used differ significantly in the distribution of fitted age of their members.
A median of age in the Sample 1, 2 and 3 is $65\pm 8$ Myr, $22\pm 2.5$ Myr, and $4\pm 1$ Myr,
respectively. A median linear size in these samples is $\sim$1330 kpc, $\sim$510 kpc, and $\sim$145
kpc, respectively. As the Samples 1 and 2 comprise the largest sources (of FRII type only) each
of them in different redshift range, and this range in the Sample 2 is twice the range of the
Sample 1 -- we compare  the $D-z$ dependence for these sources with that found in other samples
unlimited in linear size of their members (e.g. Eales 1985; Barthel \& Miley 1988). Since the
Sample 1, i.e. the sample of known high-redshift GRSs, is very small and 
consists of 12 sources only, below we also consider 12 of the largest ones from the 
Sample 2.
The relevant medians are: $D_{\rm med}=1330\pm 120$ kpc at $z_{\rm med}=0.60\pm 0.10$, and
$D_{\rm med}=760\pm 80$ kpc at $z_{\rm med}=1.35\pm 0.20$, respectively. Assuming that
$D_{\rm med}\propto (1+z)^{-n}_{\rm med}$, we find $n\approx 1.5$ which is compatible with
a value of this power $\sim$(1--2) determined in several samples of radio galaxies and quasars
(e.g. Kapahi 1989).
Therefore the above result agree with the trend observed in much more abundant samples of sources
and confirms that the largest linear size, which a source can achieve before dimming 
below the detection limit, inevitably decreases with redshift irrespective of
a cosmological evolution of the IGM (cf. Nilsson et al. 1993). 

In order to enlighten the latter problem, we pay an attention to the pressure at the head of
lobes (along the jet's axis), $p_{\rm h}$, given in column 6 of Tables 4, 5 and 6. This
pressure is determined in the DYNAGE as

\[p_{\rm h}={\cal P}_{\rm hc}(R_{\rm T})p_{\rm c}(R_{\rm T},t),\]

\noindent
where ${\cal P}_{\rm hc}$ is the pressure ratio (cf. Section 4.2), and $p_{\rm c}=(\Gamma_{\rm c}-1)u_{\rm eq}$, where  $u_{\rm eq}$ is given by Eq.\,11.
The median of this pressure, $p_{\rm h,med}$, for sources in the Sample 1 and Sample 2 is
$\sim 10^{-12.0}$ N/m$^{2}$ and $\sim 10^{-10.8}$ N/m$^{2}$, respectively, thus a (mean) pressure at the head of lobes of the sources within the redshift range $z$[1, 2] is about 16 times higher
than that for the sources within $z$[0.5, 1]. If so, it is more than twice higher than a ratio
implied from the formula giving pressure of a non-relativistic, homogeneous IGM in thermal
equilibrium, $p(z)=p_{0}(1+z)^{5}$, where $p_{0}$ is the present-day pressure (Subrahmanyan \&
Saripalli 1993). We return to this point in the next section.

\begin{figure}[t]
\begin{center}
\includegraphics[width=12cm]{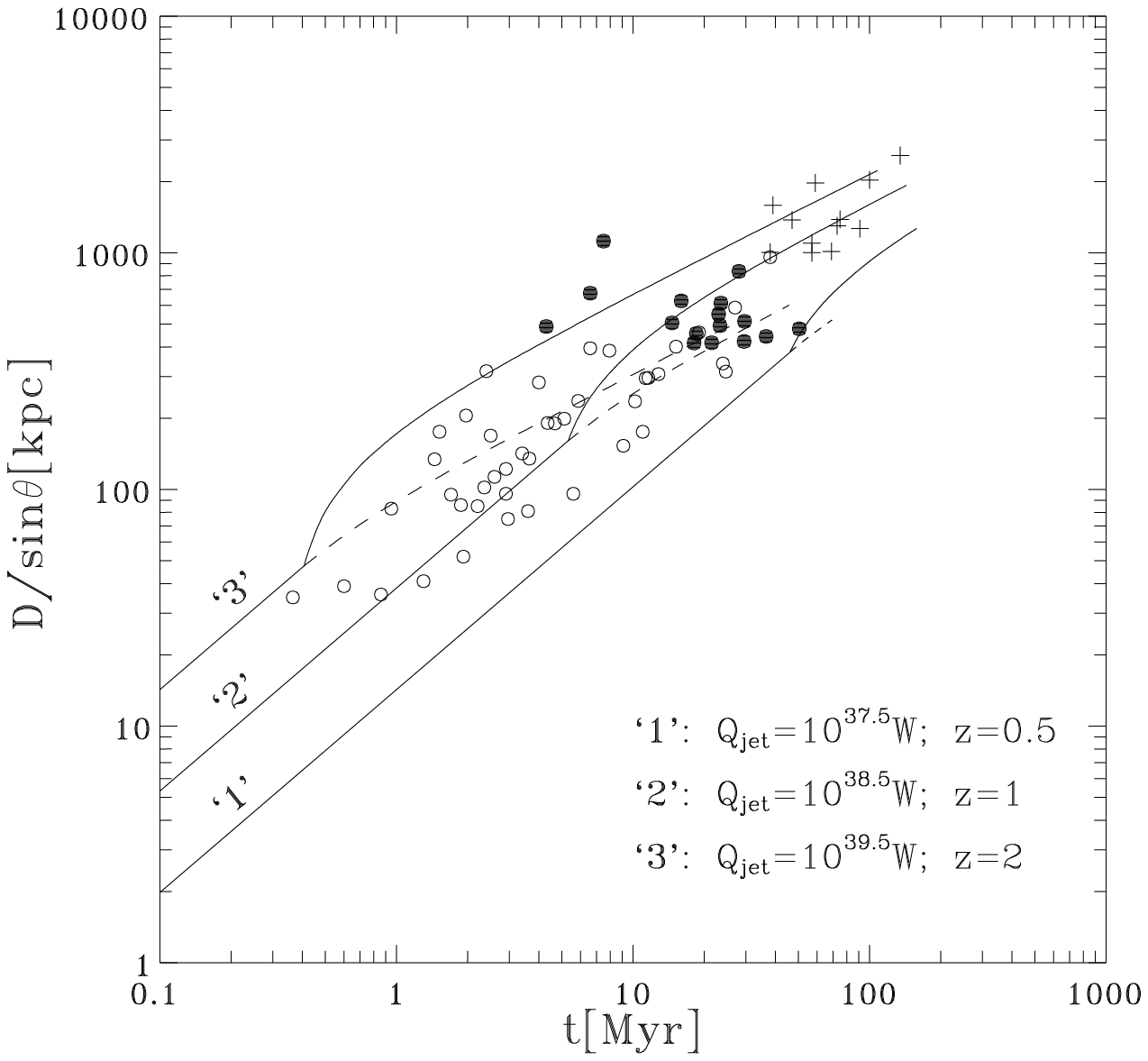}
\FigCap{ $D$ vs $t$ diagrams for three sets of the jet power and redshift. The solid lines indicate
the model predictions in the frame of Scenario A, the dashed lines -- in Scenario B. The sources
from the Samples 1, 2, and 3 are plotted with the crosses, full dots, and open circles, respectively.}
\end{center}
\end{figure}

\begin{figure}[t]
\begin{center}
\includegraphics[width=12cm]{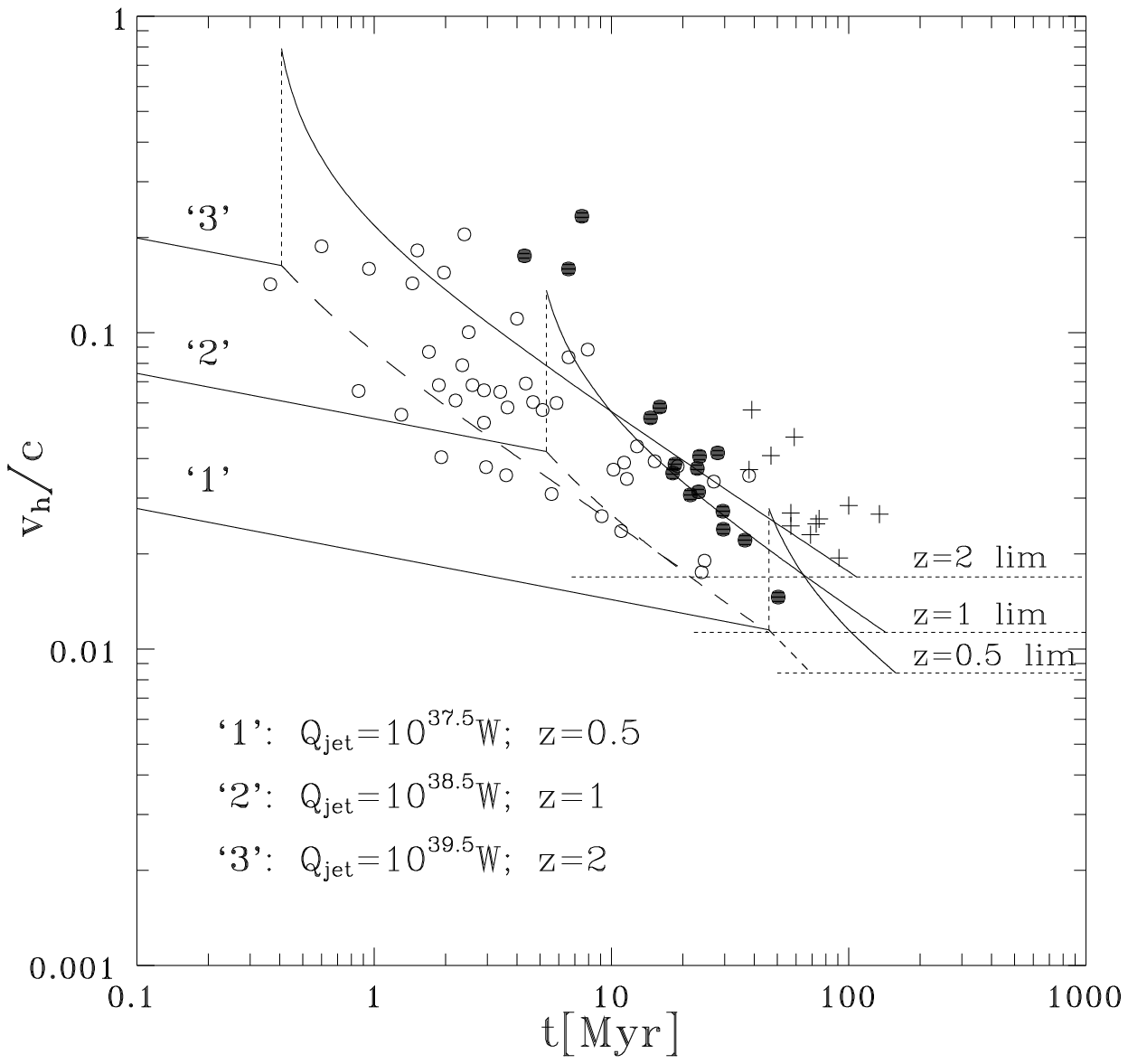}
\FigCap{$v_{\rm h}/c$ vs $t$ diagrams for the same sets of the model parameters as in Fig.\,5. The
three horizontal dotted lines indicate the lower limits for the expansion velocity determined by
the sound speed at a given redshift.}
\end{center}
\end{figure}

\begin{figure}[t]
\begin{center}
\includegraphics[width=10cm]{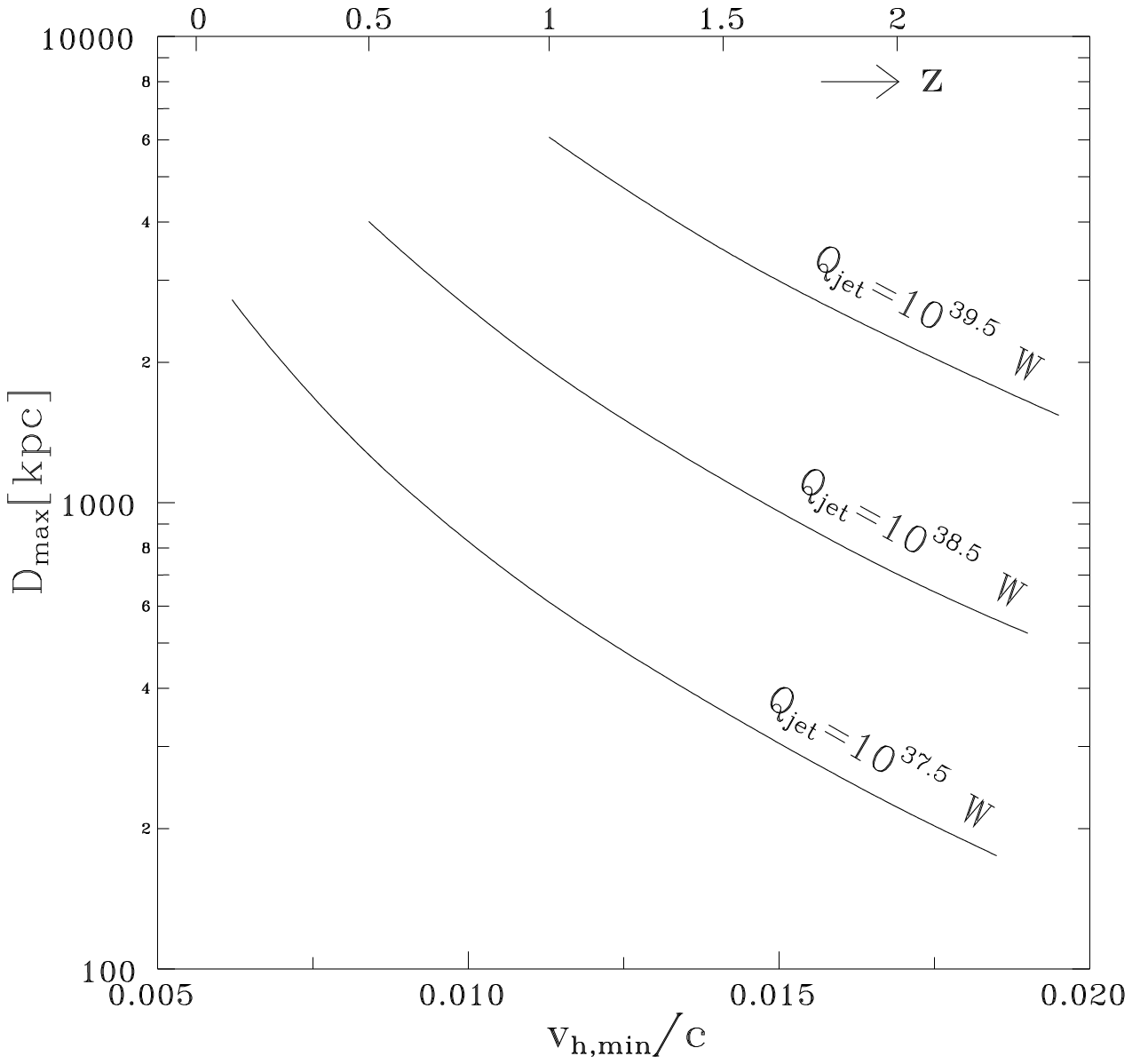}
\FigCap{ $D_{\rm max}$ vs $v_{\rm min}/c$ diagrams for three values of the jet power.}
\end{center}
\end{figure}

\begin{center}
\section{Discussion of the results and conclusions}
\end{center}

The distribution of sources on the planes ${D}/{sin\phi}$ -- $t$ and $V_{\rm h}/c$ -- $t$ 
(in Fig. 5 and 6, respectively) is more or less compatible with model's prediction, however the samples used to constrain the model may be to small to be decisive. So, much larger samples of sources with reliably determined ages would support or impair the inferences drawn as follows:

(i) Scenario B is rather excluded because both the maximum size, $D_{\rm max}\sim 600$ kpc,
 and the corresponding age, $t_{\rm max}\sim 70$ Myr allowed by the model are evidently smaller
 and younger then the observed values of those parameters for the sample sources. 
Also high-dynamics radio observations show 
no evidence for a flaring of the jets in FRII-type sources; oppositely the jets in these sources
 are rather recollimated in vicinity of the radio core.  

(ii) As expected, the inferred expansion velocity, v$_{\rm h}/c$, of all the sources used to
constrain the G-KW model are higher than the limiting sound speed marked in Fig.\.6.
This result confirms a common believe that the heads of lobes of FRII type
sources are overpressured with respect to the external gaseous medium. We note that the lowest
expansion velocities of the largest sources, i.e. these in the Samples 1 and 2, are comparable
to those of much smaller sources in the Sample 3.

(iii) The age of the three sources in Sample 2 is probably underestimated and the corresponding
expansion velocity -- overestimated (the three full dots with $t<10$ Myr and v$_{\rm h}>0.1\,c$
in Figs.\,5 and 6) due to possible relativistic effects (cf. Arshakian \& Longair 2000). All
are classified as quasars, and the two of them (J0809+2912 and J1550+3652) are highly
asymmetric in their lobes' brightness. If a probable anisotropic radiation is discerned and
the proper luminosity of the cocoon is taken into calculations, the age of these sources would
be older and the expansion velocity -- lower than the values in Table 6.    

(iv) The pressure in the diffuse lobes of the largest radio sources seems to offer a tool useful for an
estimation of the IGM pressure. If the axial pressure in lobes of the largest observed sources,
especially these in the Sample 2,  is close to equilibrium with the
IGM pressure, the inferred values of $p_{\rm h}$ would suggest even stronger density evolution
of the IGM than $\rho_{\rm IGM}\propto (1+z)^{3}$ (cf. Eqs.\,5 and 6). However, radio images
of the sources in the Sample 2 evidently indicate a presence of hot spots which, in turn,
may confirm that the hot spot regions are highly overpressured with respect to the IGM, and
the actual age of sources cannot be considered as a lifetime. Moreover, the DYNAGE fits show that
the jet powers of the most distant sources are significantly higher than that for sources of a
similar age but at low redshifts $z<0.2$. As it was shown in Section 5.2, the median 
age in the Sample 2 is $\sim 22$ Myr. An 
inspection of the compilation of over 200 radio
galaxies and quasars with the dynamical ages fitted using the DYNAGE method (the data
unpublished yet) resulted in only 7 radio galaxies of age of about 25 Myr and lying within the
redshift range $z$[0.1, 0.2]. These galaxies and their observational and dynamical  parameters
are listed in Table 7.

\begin{table}
\caption{Observational and dynamical parameters of seven FRII-type radio galaxies with $0.1^{<}_{\sim} z ^{<}_{\sim}0.2$, for which their 
dynamical age determined with the DYNAGE is 12 Myr$<$t$<$36 Myr.}
\begin{tabular}{llrcccc}
\hline
Name            & z    & $D$[kpc] & log\,$P_{1400}$ & $t$[Myr]  & log\,$Q_{\rm jet}$ & log\,$\rho_{0}$\\
  & & & [W/(Hz$\cdot$sr)] & & [W] & [kg/m$^{3}$]\\
\hline
3C332           & 0.1515 & 229   & 25.07  & 36  & 37.45  & $-$22.60\\
3C349           & 0.205  & 287   & 25.46  & 29  & 37.91  & $-$22.44\\
3C357           & 0.1664 & 296   & 25.20  & 24  & 37.77  & $-$22.96\\
3C381           & 0.1605 & 199   & 25.29  & 25  & 37.66  & $-$22.76\\
B0908+376       & 0.1047 & 229   & 24.12  & 20  & 36.50  & $-$23.59\\
B1130+339       & 0.2227 &  99   & 24.98  & 12  & 37.38  & $-$23.01\\
B1457+292       & 0.146  & 172   & 24.15  & 28  & 36.80  & $-$23.99\\
\hline
\end{tabular}
\end{table}

The median jet power, $Q_{\rm jet,med}$, in the Sample 2 (cf. Table 5) is $\sim 1.8\times 10^{39}$ W,
while that for the galaxies in Table 8 is $\sim 2.8\times 10^{37}$ W. We argue that this is very
unlikely to find so young FRII type radio sources at redshift below 0.2 and driven by jets more
powerful than $10^{38}$ W. Although this may be partly caused by the selection effect. In fact, searching
for high-redshift sources we probe a much larger space volume than the volume
corresponding to a low redshift. Probably therefore this is why the only 
low-redshift and very powerful radio galaxy known is Cygnus A with $t\approx 8$
Myr, $D=135$ kpc, and $Q_{\rm jet}\approx 1.4\times 10^{39}$ W (cf. Machalski
et al. (2007a), the above result strongly suggests that the dynamical evolution of FRII 
type radio sources at high redshifts (in earlier cosmological epochs) is different and 
faster than that of sources at low redshifts. The model constrained in this paper implies 
that giant-sized sources (with $D>1$ Mpc) can exist at redshifts as high as $\sim 2$, 
however only if their jets are powerful enough.

(v) A differentiation of the jet's (highest) power at high and low redshifts perhaps would be a strong
argument for the theoretical speculations about its dependence on the properties of a black hole
(BH) in AGN: the spin of the BH (Blandford \& Znajek 1977) and/or the accretion process and its rate
(Sikora, Stawarz \& Lasota 2007; Sikora 2009). This is very likely that the accretion of matter
onto the BH must play a dominant role in the jets' production. This led to a presumption that more
powerful jets can be due to a larger amount of material available for the accretion processes inside
AGN formed at earlier cosmological epochs.

Reasuming the above we can conclude as follows:

--- An observational quest for the largest radio sources of FRII type at high redshifts
$1<z<2$ resulted in the sample of 25 sources listed in Table 2, where 20 of 25 are found
in this paper. Because the finding sky survey used was the optical SDSS survey, all the
newly radio-identified optical objects appear to be quasars with the (projected) linear
size from $\sim$400 kpc to $\sim$1200 kpc. The above bias precluded a detection of large,
most distant and low-luminosity radio galaxies whose parent optical counterpart will likely
be of $\sim$(22--24) R mag.

--- Though the samples used to constrain the model are small, the observational data seem
to be concordant with the predictions in the frame of Scenario A. However much larger samples
of distant sources with reliably determined ages will be more decisive for the above aim.

--- The derived lowest values of the lobes' head pressure, $p_{\rm h}$, evidently higher
than the limiting sound speed at different redshifts, strongly suggest that heads of even
the largest sources at high redshifts are still overpressured with respect to the IGM.

-- An existence of giant-sized radio sources at high redshifts is possible due to extremely
high power of their jets up to $\sim 10^{40}$ W. Such the highest jet powers are likely
related to the accretion processes onto massive black holes in the central AGN, which might
be very efficient in the nuclei formed at earlier cosmological epochs.

\Section{Appendix}

Table 8 gives precise sky coordinates of the parent optical object selected from the SDSS survey and identified
with the given radio source, its (largest) angular size
measured in the FIRST images, and the 1400-MHz total flux density taken from
the NVSS catalogue.

\MakeTable{lrcccrccccl}{12.5cm}{The newly discovered FRII-type sources with 1$<$z$<$2 and
 D$>$400 kpc in Sample 2 and their observational parameters.}
{\hline
IAU name  & RA (J2000) & DEC (J2000) & LAS &  $S_{\rm 1400 MHz} $   \\
      &    [h  m  s]  &     [$^{\circ}$ ' '']  & [arcsec]  & [mJy]                                                                       
\\
\hline
J0245$+$0108        & 02 45 34.07 & $+$01 08 13.7 & 53.4 & 327   \\
J0809$+$2015        & 08 09 20.81 & $+$20 15 38.6 & 56.3 & 84.8   \\    
J0809$+$2912        & 08 09 06.22 & $+$29 12 35.5 & 129 & 304   \\
J0812$+$3031        & 08 12 40.08 & $+$30 31 09.4 & 146 & 24.5   \\
J0819$+$0549        & 08 19 41.13 & $+$05 49 42.7 & 115 & 27.4 \\
J0839$+$2928        & 08 39 51.75 & $+$29 28 18.2 & 50.5 & 113     \\
J0842$+$2147        & 08 42 39.96 & $+$21 47 10.4 & 130 & 49.2   \\
J0857$+$0906        & 08 57 48.57 & $+$09 06 48.1 & 59.1 & 130  \\
J0902$+$5707        & 09 02 07.20 & $+$57 07 37.9 & 101 & 29.8  \\
J0906$+$0832        & 09 06 49.99 & $+$08 32 55.9 & 79.8 & 78.7  \\
J0947$+$5154        & 09 47 40.01 & $+$51 54 56.8 & 58.6 & 111 \\
J0952$+$0628        & 09 52 28.46 & $+$06 28 10.5 & 65.0 & 127  \\
J1030$+$5310        & 10 30 50.91 & $+$53 10 28.6 & 100 & 56.4  \\
J1039$+$0714        & 10 39 36.67 & $+$07 14 27.4 & 58.6 & 24.7   \\
J1130$+$3628        & 11 30 26.18 & $+$36 28 36.9 & 51.6 & 45.2   \\
J1207$-$0244      & 12 07 08.02 & $-$02 44 44.2 & 54.6 & 62.1   \\
J1434$-$0123      & 14 34 10.77 & $-$01 23 41.7 & 60.6 & 119 \\
J1550$+$3652        & 15 50 02.01 & $+$36 52 16.8 & 80.4 & 47.4  \\
J1706$+$3214        & 17 06 48.07 & $+$32 14 22.9 & 54.2 & 125   \\
J2345$-$0936      & 23 45 40.45 & $-$09 36 10.2 & 60.9 & 223 \\

\hline
\multicolumn{11}{p{8cm}}{}
}

\Acknow{This project was supported by the MNiSW with funding for scientific research in years
2009--2012 under contract No. 3812/B/H03/2009/36.
Funding for the SDSS and SDSS-II has been provided by the Alfred P. Sloan Foundation, the
Participating Institutions, the National Science Foundation, the U.S. Department of Energy, the
National Aeronautics and Space Administration, the Japanese Monbukagakusho, the Max Planck
Society, and the Higher Education Funding Council for England. The SDSS Web Site is
http://www.sdss.org/.}

\vspace{2mm}
The authors are grateful to the anonymous refree for
valuable remarks and suggestions that helped to improve the paper.

\end{document}